\newcommand\ee{e E} 

\newcommand\idq{{\mathbb I}_4}
\newcommand\idd{{\mathbb I}_2}

\newcommand\pa{\partial}

\newcommand\beque{\begin{equation*}}
\newcommand\beq{\begin{equation}}
\newcommand\eeq{\end{equation}}
\newcommand\eeque{\end{equation*}}
\newcommand\beqnl{\begin{eqnarray}}
\newcommand\beqna{\begin{eqnarray*}}
\newcommand\eeqna{\end{eqnarray*}}
\newcommand\eeqnl{\end{eqnarray}}

\newcommand\RR{{\mathbb R}}

\documentclass[a4paper,10pt]{JHEP3}
\usepackage{amsmath}
\usepackage{amsfonts}
\usepackage{amsthm}
\usepackage{amsbsy}
\usepackage{amssymb}

\DeclareMathOperator{\real}{Re}
\DeclareMathOperator{\imag}{Im}

\date{\today}

\title{Pair-production of charged Dirac particles on charged Nariai and ultracold black hole manifolds}

\author{F. Belgiorno\\Dipartimento di Matematica and Dipartimento di Fisica,\\ Universit\`a di Milano, 20133
Milano, Italy 
\\ E-mail address: belgiorno@mi.infn.it}
\author{S. L. Cacciatori\\ 
Dipartimento di Fisica e Matematica, Universit\`a dell'Insubria, 22100
Como, Italy, and\\ I.N.F.N., sezione di Milano, Italy\\ 
E-mail address: sergio.cacciatori@uninsubria.it}
\author{F. Dalla Piazza\\ 
Dipartimento di Fisica e Matematica, Universit\`a dell'Insubria, 22100
Como, Italy, and\\ I.N.F.N., sezione di Milano, Italy\\
E-mail address: f.dallapiazza@uninsubria.it}

\abstract{Spontaneous loss of charge by charged black holes by means of pair-creation of charged Dirac 
particles is considered. 
We provide three examples of exact calculations for the 
spontaneous discharge process for 4D charged black holes by considering 
the process on three special non-rotating 
de Sitter black hole backgrounds, which allow to bring back the problem 
to a Kaluza-Klein reduction. Both the zeta-function approach and 
the transmission coefficient approach are taken into account.  A comparison between the 
two methods is also provided, as well as a comparison with WKB results.  
In the case of non-zero temperature of the geometric background, 
we also discuss thermal effects on the discharge process. 
}

\begin{document}


\section{Introduction}
\label{intro}

Spontaneous loss of charge by a charged black hole is a relevant topic in the
framework of quantum effects in the field of a black hole \cite{gibbons,khriplovich}.
It belongs to the framework of phenomena which are due to vacuum instability in presence of 
an external field, with consequent pair creation. Quantum-electrodynamics effects in presence of 
an external electric field have been in particular a key-topic which has been extensively 
discussed. Being our interest oriented toward an application to black hole physics, 
we limit ourselves to quote two seminal papers 
\cite{euler-heisenberg,schwinger} and, in the recent literature, Refs. \cite{gitman,kim-page,kim}. 
An effective description of pair creation phenomenon 
for static charged black holes was provided by Damour, Deruelle and 
Ruffini in a series of papers \cite{damo,deruelle,ruffini}\footnote{Also rotating black holes 
were treated therein.}. To sum up, on these backgrounds the Hamilton-Jacobi 
equations (H-J) for a classical charged particle can be easily reduced to quadrature by means of variables 
separation. In particular, the radial equation describes a one dimensional motion of a 
particle in a given potential. The H-J equation, beyond a positive energy potential, 
determines a negative energy potential which at classical level must be discarded. However, at quantum 
level, negative energy states must be included, and a quantum interpretation to this couple of potentials 
can be given. The positive energy potential determines the allowed positive energy states, whereas the 
negative energy potential determines the allowed negative energy states. The usual separation of these 
states occurring in absence of external fields is not ensured a priori, and 
there can be regions 
where an overlap of positive and negative states for the particle is allowed, i.e. the Klein paradox 
takes place. In these level-crossing regions, by means of tunneling between negative and positive states, 
pair production 
of charged particles can take place with a rate determined by the transmission probability for the particle 
to cross the forbidden region between the two potentials, and can be computed e.g. in the WKB approximation.\\
We improved this semiclassical picture in the case of anti de Sitter Reissner-Nordstr\"{o}m black holes
showing that the potentials have a direct interpretation at the quantum level 
without referring to the classical H-J equation \cite{belcaccia-ads}. Then, for the class of 
de Sitter Reissner-Nordstr\"{o}m black holes we found that level-crossing is always present, due to the 
peculiar occurrence of both a black hole event horizon and a cosmological event horizon \cite{belcaccia-rnds}, 
and we also considered a particular limit case, 
when the black hole horizon radius $r_+$ equates the cosmological horizon radius $r_c$: 
the Nariai black hole \cite{romans,bousso,mann}. 
The aforementioned class of solutions contains further limit cases, 
corresponding to the extremal cases $r_-=r_+=r_c$, which are called ultracold solutions of type I and II 
\cite{romans,mann}.  
A careful WKB analysis was also performed for the Nariai case and the ultracold ones.\\
Herein, we develop our analysis of the pair-creation process associated with the black hole 
electrostatic field, and fully exploit the fact that   
the aforementioned special backgrounds allow an exact calculation of the vacuum instability.  
As a consequence, we can provide for the first time, to our knowledge, exact results for the instability 
of 4D charged black holes. We point out that our backgrounds are of a special character:    
in all the cases the geometry involved is the one of a Cartesian product $M^{1,1}\times S^2$ where $M^{1,1}$
is a two dimensional spacetime and $S^2$ is a sphere with constant radius (there is not any 
non-constant warping factor). 
Moreover, the fluxes do not involve the sphere directions so that the sphere could be effectively considered as an
internal space and the problem can be reduced to a two dimensional effective theory by means of a Kaluza-Klein 
reduction \cite{kaluza,klein,duff,salam}. Here we will recall manifestly
the strategy of a K-K reduction in part of our analysis. 
Indeed, for all cases, the Dirac equation will be reduced to a two dimensional
Dirac equation on the $M^{1,1}$ background, with the two dimensional spinors obtained by the two dimensional reduction
of the four components Dirac spinors and the mass spectra corrected by the K-K modes. 
We also recall that in a K-K reduction 
the latter corrections are provided by the harmonic analysis of the internal space $S^2$: 
$$
\mu^2 \longrightarrow \mu_l^2=\mu^2 +\lambda_l,
$$ 
where $\mu$ is the particle mass and  
$\lambda_l$ are the eigenvalues of the Laplacian operator $-\Delta_{S^2}$ on the internal space. 
In our case we are involved with harmonic analysis for spinors, and then 
the Dirac ``angular momentum'' operator eigenvalues 
$k = \pm (j+\frac 12)\in {\mathbb{Z}}-\{0\}$ appear in $\lambda_l$.

We also take into account the fact that, both in the Nariai case and in the ultracold I one, 
the real quantum state to be considered is not the Boulware-like state corresponding to 
standard quantum vacuum, but the Hartle-Hawking state associated with the black hole temperature. 
Then, we discuss also how pair-creation due to the electrostatic field of the black hole 
superimposes to the thermal radiation effect which is present in the given backgrounds. 
We show that, in the thermal means of the Dirac field number operators, a standard thermal 
contribution appears together with a term which is still related to the aforementioned 
vacuum instability, except for a further dependence on the background temperature which 
is induced by the Hawking effect.

The plan of the paper is the following. In section \ref{vac-inst} we recall the transmission 
coefficient approach to vacuum instability, and consider also how the instability affects  
thermal states in the framework of Thermofield Dynamics. 
In section \ref{ultracoldII} we discuss the ultracold II case, 
both in the transmission coefficient approach and in the zeta function approach. 
In section \ref{ultracoldI}, an analogous analysis is carried out for the ultracold I case. 
In section \ref{nariai} the Nariai case is considered. In this case, the zeta function approach 
requires a recently developed calculation strategy \cite{caccia-zeta}, which is sketched in the present 
paper. For all the cases a comparison with WKB results is done. In section \ref{conclusions} conclusions 
appear.

\section{Vacuum instability and thermal states}
\label{vac-inst}

In this section, we first recall some aspects of the pair creation due to vacuum instability, 
in particular we focus on the so-called transmission coefficient approach to the evaluation 
of the instability, which is associated with the presence of an imaginary part of the effective 
action \cite{schwinger}, which in the following is approached also by means of $\zeta$-function techniques. 
Then we consider how instability in external fields affects thermal states, referring to Thermofield 
Dynamics approach for a general fermionic case, although our interest is in the black hole case.

\subsection{Vacuum instability in the transmission coefficient approach}

There are several ways one can deal with vacuum instability. One consists in 
Schwinger's approach \cite{schwinger,brezin,itzy}, with calculations carried out in the correct 
space-time signature, or in its $\zeta$-function variant where calculations 
are developed in Euclidean signature and then a rotation to real time is performed. 
As to vacuum instability, we adopt the $\zeta$-function variant, and a double check 
of our results is also provided by the so-called transmission coefficient approach. 
We recall shortly some aspects of the latter,  
in which, following \cite{damo,nikishov,naroz},
it is also possible to reconstruct the probability of persistence of the vacuum.
Let us introduce, for a diagonal scattering process,
\beq
n_i^{IN}= R_i n^{OUT}_i + T_i p^{OUT}_i,
\eeq
where $n_i$ stays for a negative energy mode and $p_i$ for a positive energy one.
$T_i$ is the transmission coefficient and $R_i$ is the reflection one.
Moreover, in \cite{damo} one defines
\beq
\eta_i:=|T_i|^2,
\eeq
which coincides with the mean number per unit time and unit volume 
of created particles \cite{damo}. Cf. also \cite{nikishov,gavrilov-thermal}. 
In the case of fermions, the result is 
\beq
|R_i|^2 = 1-\eta_i,
\eeq
which excludes any superradiant phenomenon for Dirac particles. As 
thoroughfully discussed in \cite{manogue}, it is crucial to consider group 
velocity for the asymptotic behavior of the solutions, in order to get a 
correct physical result. In fact, an erroneous consideration of phase velocity in place 
of group velocity could easily lead to conclude that superradiance exists also for 
the Dirac case.\\
By interpreting {\sl \`a la} Stueckelberg the scattering process, one can also obtain
\beq
n^{OUT}_i= R_i^{-1} n_i^{IN} - R_i^{-1} T_i p^{OUT}_i,
\eeq
which is interpreted as the scattering of a negative mode incident from the future
and which
is in part refracted in the past and in part reflected in the future. The new
reflection amplitude $- R_i^{-1} T_i$ is such that the reflection coefficient
\beq
|R_i^{-1} T_i|^2 = \frac{\eta_i}{1-\eta_i}
\eeq
can be interpreted as the relative probability for the creation of the pair
$n^{OUT}_i,p^{OUT}_i$. 
The absolute probability is obtained by multiplying the relative one times the
probability $p_{i,0}$ to
get zero pairs in the channel $i$, and then the probability $p_{i,n}$ of $n$ pair for fermions is
\beq
p_{i,n} = p_{i,0} \frac{\eta_i^n}{(1-\eta_i)^n} .
\eeq
The normalization condition
\beq
\sum_{n=0}^{1} p_{i,n} =1,
\eeq
for fermions leads to
\beq
p_{i,0} = 1-\eta_i.
\eeq
The persistence of the vacuum is given by
\beq
P_0 = \prod_i p_{i,0} = \exp (-2 \imag W),
\eeq
and then
\beq
2 \imag W = -\sum_i  \log (1-\eta_i) = \sum_i \sum_{k=1}^{\infty}  \frac{1}{k}
\eta_i^k
\eeq
for fermions. For bosons see e.g. \cite{damo}.

\subsection{Finite temperature effects}

In the case of ultracold I and Nariai manifolds, one has to take into account that we deal with a 
black hole manifold endowed with a non-zero temperature. As a consequence, we have to 
consider quantum instability not simply for a vacuum state which corresponds to 
the Boulware-like state of standard Schwarzschild solution, but for the state which plays 
the role of Hartle-Hawking state for the given solution. Since the discover of the 
Hawking effect, a very fine construction of the thermal state living on a finite temperature 
black hole manifold characterized by a bifurcate Killing horizon was introduced by Israel 
\cite{israel-plb} on the grounds of Unruh analysis \cite{unruh} and of the thermofield 
approach to thermal physics introduced by Takahashi and Umezawa \cite{taka-ume,umezawa-condensed,umezawa-book}. 
Israel discovered that the HH state (hence called also Hartle-Hawking-Israel state) corresponds 
to the thermal vacuum of thermofield approach with the temperature equal to the black-hole temperature, 
and that the would-be fictitious states of the thermofield approach correspond to the 
states in the left wedge of the extended solution (if one is living in the right wedge). 
In our case, one possibility is to consider an analogous construction; alternatively,  
we adopt a more ``liberal'' attitude, in the sense that we appeal to Thermofield Dynamics 
formalism and describe the equilibrium state (KMS state) without caring about the reality of the 
``would-be fictitious'' states in a ``specular wedge'' of the extended manifold. One could 
also appeal to the approach developed in \cite{moschella} (where a KMS quantization, leading to a KMS state, 
is introduced, without any doubling of the Hilbert space).\\ 
We show that the transmission coefficient approach at finite temperature 
is still a valid approach to analyze the quantum instability problem at hand. Thermofield dynamics also 
helps a straightforward generalization of quantum instability to the case where the initial (in) and the 
final (out) states are thermal states (at the same temperature) instead than vacuum ones. 
We could as well start from results given in \cite{kimleeyoon,kim-scalar} and also in 
\cite{gavrilov-thermal,gavrilov-tensor}, which analyze stability topics in quantum electrodynamics.  
In order to check if there is instability in the thermal state at the Hawking temperature, 
we adopt the following ``standard'' strategy: we calculate the mean number in the ``in'' thermal state 
of ``out'' particles in the $l$-mode, and see if there exists any deviation from a purely 
thermal distribution. Equivalently, we could evaluate the thermal mean of the number 
of ``out'' particles in the $l$-mode minus the number 
of ``in'' particles in the $l$-mode (cf. \cite{kimleeyoon}) and then see the net effect 
of a possible quantum instability. 
In what follows, our focus is to the case of black hole backgrounds with a single temperature, 
and then $\beta$ is to be meant in the black hole case as the inverse black hole temperature. 
Still, we point out that the following analysis holds true for fermions in a generic thermal state with 
inverse temperature $\beta$.\\
We refer both to \cite{damo} and to \cite{kimleeyoon}, and first we consider the 
Bogoliubov transformation for the ``diagonal'' case (we purposefully choose a notation 
which allows a straightforward comparison with \cite{kimleeyoon}; see also \cite{damo}):
\beqnl
a_l^{out} &=& \mu_l\ a_l^{in}+ \nu_l\; (b_l^{in})^{\dagger} \\ 
b_l^{out} &=& \mu_l\ b_l^{in}- \nu_l\; (a_l^{in})^{\dagger},
\eeqnl
where $l$ is a collective index specifying states in the Hilbert space, and where 
the usual CCR rules for fermions lead to 
\beq
|\mu_l|^2 + |\nu_l|^2 = 1.
\eeq
We are interested in the following operator:
\beqnl
N_l^{out} (a): &=&(a_l^{out})^{\dagger} a_l^{out} \cr
&=& |\mu_l|^2 (a_l^{in})^{\dagger} a_l^{in} + \nu^{\ast} \mu b_l^{in} a_l^{in} +
\mu^{\ast} \nu (a_l^{in})^{\dagger} (b_l^{in})^{\dagger} + 
|\nu_l|^2 \left( 1 - (b_l^{in})^{\dagger} b_l^{in} \right).
\eeqnl
For our aims, it works equally well the operator which allows to detect the net effect 
of the instability (cf. \cite{kimleeyoon}) 
\beq
\bar{N}_l^{out} (a):=(a_l^{out})^{\dagger} a_l^{out}-(a_l^{in})^{\dagger} a_l^{in}. 
\eeq
We also introduce thermal state operators, according to the standard constructions in 
thermofield dynamics \cite{umezawa-condensed,umezawa-book,das-thermal}, both for initial 
and final states. We omit ``in'' and ``out'' labels in this case, for simplicity of notation, 
and introduce the thermal state $|O(\beta)>$ and thermal state annihilation operators 
$a_l(\beta), \tilde{a}_l (\beta), b_l(\beta), \tilde{b}_l (\beta)$, which are such that 
\beq
a_l(\beta) |O(\beta)> = \tilde{a}_l (\beta) |O(\beta)> =
b_l(\beta) |O(\beta)> = \tilde{b}_l (\beta) |O(\beta)> = 0. 
\eeq
We are mainly interested in the following relations (see also \cite{khanna}): 
\beqnl
a_l &=& s^+_l a_l(\beta) + c^+_l \tilde{a}_l^{\dagger} (\beta),\\
b_l &=& s^-_l b_l(\beta) + c^-_l \tilde{b}_l^{\dagger} (\beta),
\eeqnl
with 
\beqnl
c^+_l : &=& \frac{1}{\sqrt{1+\exp [\beta (\omega-\varphi^+)]}},\\
s^+_l : &=& \frac{\exp [\frac 12 \beta (\omega-\varphi^+)]}
{\sqrt{1+\exp [\beta (\omega-\varphi^+)]}},
\eeqnl 
and the analogous ones for $b$-operators (which correspond to operators for antiparticles, i.e.
for negative frequency states; cf. \cite{umezawa-condensed}):
\beqnl
c^{-}_l : &=& \frac{1}{\sqrt{1+\exp [\beta (|\omega|+\varphi^-)]}},\\
s^{-}_l : &=& \frac{\exp [\frac 12 \beta (|\omega|+\varphi^-)]}
{\sqrt{1+\exp [\beta (|\omega|+\varphi^-)]}},
\eeqnl 
where $\varphi^+,\varphi^-$ are chemical potentials for particles and antiparticles 
respectively \cite{khanna}.\\
In the previous formula and in the following ones, for simplicity of notation we 
make a ``liberal'' use of indexes for quantum numbers, as far as unambiguous formulas arise. 
By making use of the above relations between particle operators and thermal state creation and 
annihilation operators, we easily find for particles 
\beq
<\bar{N}_l^{out}>_{\beta} = |\nu_l|^2 (1-(c^+_l)^2 -(c^-_l)^2) =|\nu_l|^2 
\frac 12 \left(\tanh [\frac 12 \beta (\omega
-\varphi^+)] + \tanh [\frac 12 \beta (|\omega|
+\varphi^-)]\right) 
\eeq
this result in the limit as $\varphi^+,\varphi^-\to 0$ 
agrees with the result displayed in \cite{kimleeyoon}. Note that this 
means that 
\beq
<N_l^{out}>_{\beta} = (c^+_l)^2+|\nu_l|^2 (1-(c^+_l)^2-(c^-_l)^2) =\frac{1}{1+\exp [\beta (\omega-
\varphi^+)]} + <\bar{N}_l^{out}>_{\beta},
\eeq
where the former term is the expected mean number of fermions in thermal 
equilibrium at the given temperature and the latter term is the net effect associated 
with the pair production induced by the presence of an electrostatic field. 
In our notation for the the Nariai case, we shall get 
\beq
<\bar{N}_k^{out}>_{\beta} = |T_k (\omega)|^2 \frac 12 \left( 
\tanh [\frac 12 \beta (\omega-\varphi^+)] + \tanh [\frac 12 \beta (|\omega|+\varphi^-)] \right)
\eeq
where $\varphi^+$ is assumed for definiteness to be the chemical potential for 
particles in the case of a positively charged black hole, $\varphi^- = \varphi^+$, 
particles are electrons (charge $-e$) and 
\beq
\varphi^+ =-e (A_0 |_{\pi}-A_0 |_{0}) = -2 e Q \frac{B}{A}.
\eeq
Formally, an analogous experession holds for the ultracold I case, which is nevertheless 
pathological (cf. sect. \ref{inst-ucI}). Furthermore, in analogy with Page's analysis 
in \cite{page}, one can expect that charged particles are efficiently emitted with thermal 
spectrum only for small black hole masses, and then that above the corresponding threshold 
charged particles are emitted only because of the electrodynamic instability (i.e. as if 
they were emitted in vacuum).\\
Moreover, 
we have to take into account that both in the ultracold I and in the 
Nariai case we work with dimensionless (rescaled) variables and then we get
$\beta = 2 \pi$. Note also that
\beq
<0\; in | N_k^{out} |0\; in> = |T_k (\omega)|^2.
\eeq
It is also interesting to evaluate the following quantity:
\beq
( \Delta N_l )^2 :=  <(N_l^{out})^2>_{\beta} - {<N_l^{out}>}^2_{\beta}. 
\eeq
As to the operator $(N_l^{out})^2$, being $(N_l^{out})^2  =  N_l^{out}$, one finds
\beqnl
( \Delta N_l )^2  &=& <N_l^{out}>_{\beta} (1-<N_l^{out}>_{\beta})\cr 
&=& (c^+_l)^2 (1-(c^+_l)^2) + |\nu_l|^2 (1- |\nu_l|^2) 
-|\nu_l|^2 (1- |\nu_l|^2) [(c^+_l)^2 +(c^-_l)^2] \cr
&+&|\nu_l|^4 [(c^+_l)^2 +(c^-_l)^2] (1-(c^+_l)^2 -(c^-_l)^2) 
-2 |\nu_l|^2 (c^+_l)^2 (1-(c^+_l)^2 -(c^-_l)^2).
\eeqnl
In a stable equilibrium situation, i.e. in our case in absence of electrostatic charge, 
one would obtain 
\beq
( \Delta N_l )^2 = (c^+_l)^2 (1-(c^+_l)^2),
\eeq
which corresponds to the first contribution displayed above. Moreover, it is interesting to 
point out that 
\beq
<0\; in | (N_l^{out})^2|0\; in>- (<0\; in | N_l^{out} |0\; in>)^2 = |\nu_l|^2 (1- |\nu_l|^2), 
\eeq
which amounts to the temperature-independent contribution in the previous formula, where 
also a third contribution, which ``mixes'' the pair creation effect to the thermality of the 
background, occurs.


\section{Ultracold II case}
\label{ultracoldII}

The ultracold II metric is obtained from the Reissner-Nordstr\"om-de Sitter one
in the limit of coincidence of the Cauchy horizon, of the black hole event horizon
and of the cosmological event horizon:  $r_-=r_+=r_c$. See \cite{romans,mann}. In particular,
the metric we are interested in is
\beq
ds^2 = -d t^2 + dx^2+ \frac{1}{2\Lambda} (d\theta^2+ \sin^2 (\theta) d\phi^2),
\label{ultracold-II}
\eeq
with $x\in \RR$ and $t\in \RR$. Then the $(t,x)$-part of the metric is a 2D Minkowski space, 
to which a spherical part is warped with a constant warping factor. 
One gets $\Gamma_{33}^2 = -\sin (\theta) \cos (\theta),
\Gamma_{23}^3 = \cot (\theta)$. 
The electromagnetic field strength is $F=-\sqrt{\Lambda}  dt\wedge dx$, and we
can choose $A_0 = \sqrt{\Lambda} x$ and $A_j=0$, $j=1,2,3$. It is also useful to
define $E := \sqrt{\Lambda}$, which represents the intensity of the electrostatic
field on the given manifold. We note that it is uniform, and then one expects naively
to retrieve at least some features of Schwinger's result.

\subsection{The transmission coefficient approach}

Let us consider the reduced Hamiltonian which can be obtained by variable separation 
as in \cite{belcaccia-rnds}: starting from the full Dirac equation $(\slash \!\!\!\! D-\mu)\Psi=0$, 
variable separation 
leads to the following reduced Hamiltonian:
\beq
h_k = \left[
\begin{array}{cc}
-e \sqrt{\Lambda} x -\mu
& \pa_{x} +\sqrt{2 \Lambda} k \cr
-\pa_{x} + \sqrt{2 \Lambda} k
&-e \sqrt{\Lambda} x +\mu
\end{array}
\right].
\label{hk-ucII}
\eeq
$k=\pm (j+\frac 12)\in {\mathbb Z}-\{0\}$ is the angular part eigenvalue, $\mu$ and $e$ 
are the fermion mass and charge respectively. 
Then one obtains the following equation for 
$\Phi=e^{-i\omega \psi}\Psi=\left( \begin{array}{c} 
\phi_1 \cr 
\phi_2 \end{array} \right)$:
\beq
\left[-(eE x+\omega) {\mathbb{I}}_2 + i \sigma_2 \partial_x + \sqrt{2\Lambda} k \sigma_1 -\mu \sigma_3 
\right] \Phi =0,
\eeq
where $\sigma_i$ are Pauli matrices and ${\mathbb{I}}_2$ is the $2\times 2$ identity matrix. 
Let us take the
unitary transformation $\Phi=e^{-i\frac \pi4 \sigma_1} \xi$ so that
$\sigma_i \mapsto e^{-i\frac \pi4 \sigma_1}  \sigma_1 e^{i\frac \pi4 \sigma_1}$ and in particular
\beq
(\mathbb{I}_2,\sigma_1,\sigma_2,\sigma_3) \mapsto
(\mathbb{I}_2,\sigma_1,\sigma_3,-\sigma_2).
\label{unitary}
\eeq
Then one obtains
\beq
\left[-(eE x+\omega) {\mathbb{I}}_2 + i \sigma_3 \partial_x + \sqrt{2\Lambda} k \sigma_1 +\mu \sigma_2 
\right] \xi =0,
\eeq
which amounts to the following couple of differential equations
\beqnl
&& - (eE x+\omega) \xi_1 + i \partial_x \xi_1 + (\sqrt{2\Lambda} k -i\mu) \xi_2 =0,\cr
&& (\sqrt{2\Lambda} k +i\mu) \xi_1 - (eE x+\omega) \xi_2 - i \partial_x \xi_2   =0.
\eeqnl
Then we can get
\beq
\xi_2 = \frac{1}{\sqrt{2\Lambda} k -i\mu} \left[ (eE x+\omega) \xi_1 - i \partial_x \xi_1\right],
\eeq
and the following equation for $\xi_1$ is obtained:
\beq
\frac{d^2 \xi_1}{dx^2} + \left[ (eE x +\omega)^2 +i eE -\mu^2_k \right]\xi_1=0,
\eeq
where $\mu_k^2:=\mu^2+2\Lambda k^2$ is the effective mass corrected by K-K modes. We can define
\beq
y= \sqrt{\frac{2}{eE}} (eE x +\omega),
\eeq
and then we obtain the following equation 
\beq
\frac{d^2 \xi_1}{dy^2} + \left[ \frac 14 y^2 -\left(\frac{\mu^2_k}{2 eE} -\frac i2 \right)\right]\xi_1=0,
\eeq
whose solutions are parabolic cylinder functions \cite{abramowitz}. 
The calculation is completely analogous to the
one performed in \cite{damo}, and as in \cite{damo} one can easily show that the
transmission coefficient satisfies
\beq
|T_k (\omega)|^2 = \exp \left( -\pi \frac{\mu_k^2}{e E} \right).
\eeq
The latter expression corresponds to the mean number per unit time and unit 
volume of created pairs and 
coincides with the WKB approximation for the same coefficient
\cite{belcaccia-rnds}. 
This means that the WKB approximation is actually exact for the given case. This 
result is expected, being our case easily realized as a completely analogous to 
the standard case except for the compact character of the 2D transverse space. 
By adopting the strategy in
\cite{damo},
or also the one in \cite{khriplovich} for determining the factor preceding the
exponential terms, one obtains the following imaginary 
part of the effective action: 
\beq
\imag W=-\frac 12 \frac{e E S}{2\pi} \sum_{k\in {\mathbb Z}-\{0\}} g(k) 
\log (1 - \exp \left (-\frac{\pi \mu_k^2}{e E} \right) ),
\label{eff-action}
\eeq
where $g(k)=2(2 l+1)$ is the degeneracy factor.\\ 
A comparison with the standard flat space-time case \cite{schwinger,lin} shows that the
structure of the ultracold II manifold, which is a product of a 2D flat-spacetime
times a 2D sphere, yields to the substitution of the (Gaussian) integral over the
transverse dimensions with an infinite sum over $k$, still to be valued. We recall
that in the flat space-time case, one finds for the 4D density of the imaginary part of the 
effective action 
\beq
w = 2 \frac{(e E)^2}{8 \pi^3} \sum_{n=1}^{\infty}
\frac{1}{n^2} \exp \left (-\frac{\pi \mu^2}{e E} n \right),
\eeq
which, by avoiding to perform the integration on the transverse variables
$\vec{p}_{\perp}$, becomes
\beqna
w &=& 2 \frac{(e E)}{8 \pi^3} \int d^2 \vec{p}_{\perp} \sum_{n=1}^{\infty}
\frac{1}{n} \exp \left (-\frac{\pi (\mu^2+\vec{p}_{\perp}^2)}{e E} n \right)\cr
&=& -2 \frac{|e E|}{8 \pi^3}   \int d^2 \vec{p}_{\perp}
\log  \left(1- \exp \left (-\frac{\pi (\mu^2+\vec{p}_{\perp}^2)}{e E} \right) \right),
\eeqna
which is of course strictly analogous to (\ref{eff-action}).

\subsection{The $\zeta$-function approach}

We tackle the problem also by the zeta function method. This will
be useful later to compare with the more difficult Nariai case. The Euclidean 
formulation is used. For convenience, we recall that the spectral zeta function associated to an operator $H$ 
having eigenvalues $\lambda_n$ with degeneration $d_n$, is defined by
$$
\zeta_H (s)=\sum_{n=0}^\infty \frac {d_n}{\lambda_n^s}=\frac 1{\Gamma(s)} \int_{0}^\infty x^{s-1} {\rm Tr} e^{-Hx} dx,
$$
that is substantially the Mellin transform of the kernel\footnote{We assumed here that the spectrum is strictly
positive, but in general this can be weakened by complex analyticity techniques.} $K_H(x)={\rm Tr} e^{-Hx}$.
The point is that $-\log \det H= \frac d{ds} \zeta_H (0)$ defines the Euclidean effective action. After turning back to the Lorentzian
signature, the instability is measured by the imaginary part of the effective action, which must thus compared with the vacuum persistence
computed with the previous method.
To this aim, let us consider the Dirac Euclidean equation for the ultracold II background.
The spectrum of the Dirac operator, as well-known, is neither positive definite nor semi-bounded. 
To overcome this problem, one relates the zeta function of the Dirac operator $\slash \!\!\!\! D$ to the zeta function of its square, and for the sake of completeness we recall some key points.  
To get the right definition, it is convenient to start with an heuristic reasoning. 
Let $\Psi$ an eigenfunction of the operator $\slash \!\!\!\! D-\mu$:
$$
(\slash \!\!\!\! D-\mu)\Psi=\lambda \Psi.
$$
Then we have $(\lambda_{\pm}+\mu)=\pm \sqrt {\slash \!\!\!\! D^2}$ and thus we can formally write
$$
\log (\det ({\slash \!\!\!\! D-\mu}))=
\frac 12 \log (\det(-\mu+ \sqrt {\slash \!\!\!\! D^2}))+
\frac 12 \log (\det(-\mu- \sqrt {\slash \!\!\!\! D^2}))=
\frac 12 \log (\det(\mu^2-\slash \!\!\!\! D^2)).
$$
The factor $1/2$ arises from the double degeneration of each eigenvalue \footnote{If
$(\slash \!\!\!\! D-\mu)\Psi_\pm=\lambda_\pm \Psi_{\pm}$ then, for example,
$(-\mu+ \sqrt {\slash \!\!\!\! D^2})\Psi_\pm=\lambda_+ \Psi_\pm$.}. Thus, it is convenient to define
\beq
-\log (\det ({\slash \!\!\!\! D-\mu}))=\frac{1}{2}\zeta'_{\mu^2-\slash \!\!\!\! D^2} (0),
\eeq
and then for the Euclidean effective action we get
\beq
W = \frac 12 \zeta'_{\mu^2-\slash \!\!\!\! D^2} (0).
\label{effaction}
\eeq
The simplest way to proceed is to compute the eigenvalues of $-\slash \!\!\!\! D^2$, 
and next to add the mass square $\mu^2$.\\

We fully exploit the K-K reduction in order to perform the 
$\zeta$-function calculation. We first note that for the 4D Dirac operator we have
\beq
\slash \!\!\!\! D =: \slash \!\!\!\! E + \slash \!\!\!\! F, 
\eeq
where $\slash \!\!\!\! E$ depends only on variables for the first 2D factor 
of the metric and analogously $\slash \!\!\!\! F$ depends only on spherical variables 
of the last 2-sphere factor. When one considers $-D^2$, it is easy to realize that 
one obtains
\beq
-D^2 = -E^2 - F^2, 
\eeq
and then the eigenvalue $\lambda^2$ of $-D^2$ is the sum of the eigenvalue $w^2$ of 
$-E^2$ and of the eigenvalue $b^2 k^2$ of $- F^2$ ($b$ is a constant related to the 
radius of the 2-sphere factor and $k$ is the usual eigenvalue for the angular momentum 
operator $K$, which is such that $-F^2 = b^2 K^2$):
\beq
\lambda^2 = w^2 + b^2 k^2. 
\eeq 
In the ultracold cases, one finds $b^2 = 2\Lambda$; in the Nariai case, one has $b^2 = B$. 
Eigenfunctions for $-D^2$ (and then also for $-D^2+\mu^2$) are tensor products of eigenfunctions 
of $-E^2$ and of eigenfunctions of $- F^2$. As a consequence, degeneracy can be read from the 
aforementioned tensor product structure.

In what follows, we define $\tilde{\gamma}_{\mu}$, $\mu=0,1,2,3$, as the Euclidean version 
of the usual gamma matrices which are such that 
$\{\tilde{\gamma}_{\mu},\tilde{\gamma}_{\mu}\}=2\delta_{\mu,\nu}$.\\ 

In the present case, we obtain 
the following axpression for $E$: 
\beq
\slash \!\!\!\! E = \tilde{\gamma}_0 (\partial_t - i e E x)+\tilde{\gamma}_1 \partial_x,
\eeq
and then 
\beq
E^2 = (\partial_t - i e E x)^2 + \partial_x^2 + i e E \tilde{\gamma}_0 \tilde{\gamma}_1. 
\eeq
The first two terms are meant to be multiplied by 4D identity $\idq$. Being 
$i e E \tilde{\gamma}_0 \tilde{\gamma}_1 = e E \sigma_2 \otimes \idd$, and $\idq = \idd \otimes\idd$, 
the eigenvalue equation 
\beq
-E^2 \psi = w^2 \psi
\eeq
is equivalent to the following reduced 2D equation 
\beq
\left[-(\partial_t - i e E x)^2 \idd - \partial_x^2 \idd - eE \sigma_2 \right]\xi = w^2 \xi;  
\eeq
a further unitary rotation such that $\sigma_2 \mapsto \sigma_3$ 
carries the problem in a ``diagonal'' form
\beq
(- \partial_x^2 -(\partial_t - i e E x)^2 \mp eE )\eta_{\pm} = w^2 \eta_{\pm}, 
\eeq
where $\eta_{\pm}$ are components of the 2D vector eigenfunction of the above eigenvalue problem. 
Variable separation 
\beq
\eta_{\pm}(t,x) = \exp (-i\omega t) \zeta_{\pm} (x)
\eeq
leads to the following differential equations: 
\beq
\partial_x^2 \zeta_{\pm} + \left[ \pm \ee-(\omega + \ee x)^2 + w^2 \right] \zeta_{\pm}=0; 
\eeq
defining the new variable
$z = \frac{(\omega + \ee x)}{\sqrt{\ee}}$, we get 
\beq
\frac{d^2 \zeta_{\pm} (z)}{dz^2} + \left[\pm 1 + \frac{w^2}{\ee} -z^2 \right] \zeta_{\pm} (z)=0,
\eeq
which is easily realized to be the equation one obtains for the standard simple harmonic
oscillator. The quantization conditions are
\beq
2 n +1 =\pm 1-\frac{w^2}{\ee},
\eeq
which lead to
\beqnl
w^2_+ &=& 2 n \ee, \\
w^2_- &=& 2 (n+1) \ee,
\eeqnl
and then one can write
\beq
w^2 = 2 n \ee,
\eeq
by taking into account that $n=0$ has degeneracy which is one half the degeneracy of 
$n>0$ (cf. also the 2D case in \cite{blissepf}).\\
As a consequence, the eigenvalues for $-D^2+\mu^2$ are
\beq
\lambda^2 =2 n \ee + \mu^2_k,
\label{eigenvalue-ucII}
\eeq
where $\mu^2_k = 2 \Lambda k^2 +\mu^2$. Notice that the eigenvalues do not depend on $\omega$. An overall   
degeneracy factor $d$ must be determined. This can be obtained by comparing the behavior of the kernel 
$K(x)$ in $x=0$ to the universal coefficients provided by the heat kernel theorems. We have 
\begin{eqnarray*}
K(y)=\sum_k g(k)\exp(- \mu^2_k y) 
\left[ d \left( 2 \sum_{n=0}^{\infty} \exp( - 2 n \ee y) -1 \right) \right].
\end{eqnarray*}
The part in square bracket is the heat kernel for the 2D operator $-E^2$, and in the limit 
as $y\to 0$ it holds $\left[ \cdots \right]\sim 2 d (2\ee y)^{-1}$.  
As it must be equal to $2 S(4\pi y)^{-1}$ (see \cite{gilkey}, p. 368), 
where $S$ is the volume of the 2D space, we 
get $d = \frac{e E S}{2\pi}$.
We now define
\beq
\zeta (s) =: \sum_k g(k) \zeta_k (s),
\eeq
with\footnote{We introduce a factor $\frac 12$ in view of (\ref{effaction}).} 
\beqnl
\frac 12 \zeta_k (s)&=&\frac{d}{\Gamma (s)} \int_0^{\infty} dt\; t^{s-1}  \sum_{n=0}^{\infty} 
\exp( - 2 n \ee t - \mu^2_k t)- \frac 12 \frac{d}{\Gamma (s)} \int_0^{\infty} dt\; t^{s-1} 
\exp( - \mu^2_k t)
\cr
&=& \frac{e E S}{2\pi}\left[ (2\ee)^{-s} \zeta_H (s, \frac{\mu_k^2}{2\ee})-\frac{1}{\mu_k^{2s}}\right].
\eeqnl
By putting $\ee \mapsto i \ee$ we finally obtain
\beqnl
\imag \frac 12 \zeta'_k (0) &=& \frac{e E S}{2\pi}
\left[-\frac 12 \log (2\ee) + \frac{\pi \mu_k^2}{4 \ee}-\frac 12 \log (2\pi)
+\real \log \Gamma \left(-i \frac{\mu_k^2}{2\ee}\right)+\frac 12 \log (\mu_k^2)\right]\cr
&=& \frac{e E S}{2\pi} \left[ -\frac{1}{2} 
\log \left(1-\exp \left(-\frac{\pi\mu_k^2}{\ee}\right)\right)\right].
\eeqnl
The volume factor appears here because of integration over the whole spacetime is included. 
This result agrees with the previous one. 

\section{The ultracold I case}
\label{ultracoldI}

A second extremal limit of the Nariai background is given by the type I ultracold solution 
when  $r_-=r_+=r_c$. 
The metric is \cite{mann}
\beq
ds^2 = -\chi^2 d \psi^2 + d\chi^2+ \frac{1}{2\Lambda} (d\theta^2+
\sin^2 (\theta) d\phi^2),
\label{ultracold-I}
\eeq
with $\chi\in (0,\infty)$ and $\psi\in \RR$, and the electromagnetic field strength is 
$F=\sqrt{\Lambda} \chi d\chi\wedge d\psi$. The spacetime presents the structure of a 
2D Rindler manifold times a two dimensional sphere (with a constant warping factor). 
One gets $\Gamma_{01}^0 = \frac{1}{\chi},
\Gamma_{00}^1 = \chi, \Gamma_{33}^2 = -\sin (\theta) \cos (\theta),
\Gamma_{23}^3 = \cot (\theta)$. 
We can choose $A_0 = \frac{\sqrt{\Lambda}}{2} \chi^2$ and $A_j=0$, $j=1,2,3$ as potential. 
The situation is now a little bit more tricky, but we still are able to compare the 
transmission coefficient approach with the zeta function method. We expect to find out results 
which are analogous to some extent to the ones obtained in \cite{gabriel-spindel} for the 
case of a charged scalar field in a 2D Rindler spacetime with external electrostatic 
field. 

\subsection{The transmission coefficient approach}

In \cite{belcaccia-rnds} we have obtained that variable separation 
allows to obtain the following reduced Hamiltonian:
\beq
h_k = \left[
\begin{array}{cc}
-\frac{e \sqrt{\Lambda}}{2} \chi^2  -\mu \chi \hphantom{...}
& \chi \pa_{\chi} +\sqrt{2 \Lambda} k \chi\cr
-\chi \pa_{\chi} + \sqrt{2 \Lambda} k \chi \hphantom{...}
&-\frac{e \sqrt{\Lambda}}{2} \chi^2  +\mu \chi
\end{array}
\right].
\label{hk-ucI}
\eeq
Again $\mu,e$ are the mass and the charge of the fermion. We also introduce $E:=\sqrt{\Lambda}$. 

Using the coordinate $t=\chi^2/2$ the Dirac equation takes the (Hamiltonian) form \cite{belcaccia-rnds}
$$
[(eEt+\omega) I_2 +\mu\sqrt {2t} \sigma_3 -i2t\sigma_2 \partial_t -2 \sqrt{2\Lambda} k 
\sigma_1]\zeta= 0.
$$
Rotating to $\xi$ by the $\sigma_1$ Pauli matrix as before: $\zeta=e^{-i\frac \pi4 \sigma_1} \xi$ , we get
\begin{eqnarray*}
&& [2t\partial_t+i(eEt+\omega)]\xi_1-(\mu+i\sqrt{2\Lambda} k)\sqrt{2t} \xi_2=0,\\
&& [2t\partial_t-i(eEt+\omega)]\xi_2-(\mu-i\sqrt{2\Lambda} k)\sqrt{2t} \xi_1=0.
\end{eqnarray*}
This is equivalent to the system
\begin{eqnarray}
&& (\mu+i\sqrt{2\Lambda} k)\sqrt{2t} \xi_2=[2t\partial_t+i(eEt+\omega)]\xi_1 \label{xi2daxi1} \\
&& 0=t\xi_1''+\frac 12 \xi_1'-\frac 1{4t} [2t\mu_k^2-(eEt+\omega)^2-i(eEt-\omega)]\xi_1.
\end{eqnarray}
Set $\xi_1(t)=t^{-\frac i2 \omega} e^{-\frac i2 eEt} F(t)$. Then, $F$ must satisfy the confluent hypergeometric 
differential equation
$$
0=tF''+(\frac 12 -i\omega -ieEt)F'-\frac {\mu_k^2}2 F,
$$
which has general solution in terms of Kummer functions $\Phi (a;c;z)$:  
$$
F=\alpha \Phi(\frac {\mu_k^2}{2ieE};  \frac 12 -i\omega; ieEt)+\beta t^{\frac 12 +i
\omega}
\Phi(\frac 12 +i\omega+\frac {\mu_k^2}{2ieE};  \frac 32 +i\omega; ieEt).
$$
Using the Kummer relation
$$
\Phi(a;c;z)=e^z \Phi(c-a;c;-z)
$$
we get
\beq\label{Ixi1}
\xi_1(t)=\alpha e^{-\frac i2 eEt} t^{-\frac i2 \omega} \Phi(\frac {\mu_k^2}{2ieE};
\frac 12 -i\omega; ieEt)
+\beta e^{\frac i2 eEt} t^{\frac 12 +\frac i2 \omega} \Phi(1-\frac {\mu_k^2}{2ieE};
\frac 32 +i\omega; -ieEt).
\eeq
To compute $\xi_2$ we use (\ref{xi2daxi1}) and
$$
\Phi'(a;c;z)=\frac ac \Phi(a+1;c+1;z).
$$
We get
\begin{eqnarray}
&& \xi_2(t)=\frac {ieE \alpha}{\mu+i\sqrt 2 Ek} \frac {\mu_k^2}{ieE(1-2i\omega)}
e^{-\frac i2 eEt} t^{-\frac i2 \omega} \sqrt {2t}
\Phi(\frac {\mu_k^2}{2ieE} +1; \frac 32 -i\omega; ieEt)\cr
&& \qquad\ -\frac {ieE \beta}{\mu+i\sqrt 2 Ek} \frac {2ieE-\mu_k^2}{ieE(3+2i\omega)}
e^{\frac i2 eEt} t^{\frac 12 +\frac i2 \omega} \sqrt {2t}
\Phi(2-\frac {\mu_k^2}{2ieE}; \frac 52 +i\omega; -ieEt)\cr
&& \qquad\ +\frac {1+2i(\omega+eEt)}{\sqrt {2t} (\mu+i\sqrt 2 Ek)} \beta e^{\frac i2
eEt} t^{\frac 12 +\frac i2 \omega}
\Phi (1-\frac {\mu_k^2}{2ieE}; \frac 32 +i\omega; -ieEt).
\label{Ixi2}
\end{eqnarray}
The asymptotic behaviors of these solutions are, for $t\approx 0$
\begin{eqnarray}
&& \xi_1 \approx \alpha e^{-\frac i2 eE t} t^{-\frac i2 \omega}, \label{c} \\
&& \xi_2 \approx  \frac {\sqrt 2}{\mu+i\sqrt 2 Ek} \ \beta (i\omega +\frac 12)
e^{\frac i2 eE t} t^{\frac i2 \omega},\label{d}
\end{eqnarray}
whereas for $t\approx +\infty$
\begin{align}
& \xi_1\approx e^{-\frac i2 eE t} t^{-\frac i2 \omega-\frac {\mu_k^2}{2ieE}}
\left[ \alpha (-ie E)^{-\frac {\mu_k^2}{2ieE}}  \frac {\Gamma(\frac 12
-i\omega)}{\Gamma (\frac 12 -i\omega -\frac {\mu_k^2}{2ieE})}
+\beta (-ieE)^{-(\frac 12 +i\omega +\frac {\mu_k^2}{2ieE})}
\frac {\Gamma(\frac 32 +i\omega)}{\Gamma (1-\frac {\mu_k^2}{2ieE})} \right] \label{a}\\
& \xi_2\approx e^{\frac i2 eE t} t^{\frac i2 \omega+\frac {\mu_k^2}{2ieE}} \frac
{\sqrt 2}{\mu+i\sqrt 2 Ek}\
\left[ \alpha (ie E)^{\frac {\mu_k^2}{2ieE}+i\omega +\frac 12}  \frac {\Gamma(\frac
12 -i\omega)}{\Gamma (\frac {\mu_k^2}{2ieE})}
+\beta (ieE)^{\frac {\mu_k^2}{2ieE}} \frac {\Gamma(\frac 32 +i\omega)}{\Gamma (\frac
12 +i\omega +\frac {\mu_k^2}{2ieE})}\right].\label{b}
\end{align}
It is useful to introduce a new variable $x:=\log (\chi)$. 
Using group velocity, (\ref{a}) and (\ref{b}) represent ingoing and outgoing
particles respectively at $x\approx \infty$. Similarly, (\ref{c})
and (\ref{d}) are the outgoing and ingoing particle respectively at $x\approx
-\infty$. In this situation, we can compute the transmission and reflection
coefficients. Rewriting (\ref{c}) and (\ref{d}) as
$$
\xi_1\approx  A e^{-i\phi(x)} , \qquad\ \xi_2\approx B e^{i\phi(x)}
$$
and (\ref{a}) and (\ref{b}) as
$$
\xi_1\approx  C e^{-i\phi(x)} , \qquad\ \xi_2\approx D e^{i\phi(x)},
$$
with $\phi(x)=\left(\omega+\frac {\mu_k^2}{eE}\right)x +\frac 12 eE e^{2x}$, then we
must put $C=0$ so that $R=A/B$ and $T=D/B$. We get
\begin{eqnarray}
&& |R_k (\omega)|^2 =\frac {e^{-\pi \omega} \sinh \frac {\pi \mu_k^2}{2eE}}{\cosh
[\pi(\omega-\frac {\mu_k^2}{2eE})]},\\
&& |T_k (\omega)|^2 =\frac {e^{-\frac {\pi \mu_k^2}{2eE}} \cosh (\pi \omega)}{\cosh
[\pi(\omega-\frac {\mu_k^2}{2eE})]}.
\end{eqnarray}
As a check we note that $|T_k (\omega)|^2+|R_k (\omega)|^2=1$, and $|T_k (\omega)|^2$ gives the mean 
number of created pairs per unit time and unit volume.\\

A comparison with the result obtained in the WKB approximation in \cite{belcaccia-rnds} 
is also in order. We recall that WKB gives for $|T|^2$ the same result as in the 
ultracold II case. Then, the dependence on $\omega$ is missing. We can relate this 
result with the exact one as follows: one considers $eE$ fixed and $\omega\to -\infty$. 
It is easily seen that the above exact result converges to the one of the WKB approximation 
in this limit.\\


We now calculate the imaginary part of the effective action. We write again 
\beq
\imag W := \sum_k g(k) W_k.
\label{im-wk}
\eeq
First of all, we note that there is an important difference with respect to the type II case. 
Here, the level-crossing region, i.e. the region where particle and antiparticle states overlap 
(cf. e.g. \cite{damo}), is no more
the whole energy range but, assuming for definiteness $\ee>0$, is determined by $\omega\leq 0$. 
As pair production is expected to
happen only in the level-crossing region for the case $\ee>0$, we must calculate
\beq
W_k = -\frac 12 \sum_{\omega} \log (1-|T_k (\omega)|^2),
\eeq
for $\omega\leq 0$. We have $\sum_{\omega} \mapsto \frac{{\cal T}}{2\pi} \int d\omega$ (cf. 
\cite{bmpp-spindel,gabriel-spindel}), where ${\cal T}$ stays for a finite time interval, 
and it is easy to show that
\beq
\log (1-|T_i|^2) = \log \left(1 - \exp \left(- \frac{\pi \mu_k^2}{\ee} \right) \right)
- \log \left( 1 + \exp \left( 2\pi \omega -  \frac{\pi \mu_k^2}{\ee} \right) \right).
\eeq
We have to evaluate the integral
\beq
\int_{-\infty}^0 d\omega
\log \left( 1 + \exp \left( 2\pi \omega -  \frac{\pi \mu_k^2}{\ee} \right) \right)
= -\frac{1}{2\pi} {\mathrm{Li}}_2 \left(
-\exp \left( -  \frac{\pi \mu_k^2}{\ee} \right) \right), 
\eeq
and then we get
\beq
W_k = -\frac 12 \frac{{\cal T}}{2\pi} \left[\left(\int_{-\infty}^0 d\omega \right)
\log \left(1 - \exp \left(- \frac{\pi \mu_k^2}{\ee} \right) \right)
+ \frac{1}{2\pi} {\mathrm{Li}}_2 \left(
-\exp \left( -  \frac{\pi \mu_k^2}{\ee} \right) \right) \right].
\label{wk-ucI}
\eeq
The factor $\frac{{\cal T}}{2\pi} \left(\int_{-\infty}^0 d\omega \right)$ amounts to a degeneracy factor  which we can 
evaluate following \cite{bmpp-spindel}. As explained in the introduction the geometry of the ultracold I manifold is of 
the form of a Cartesian product $M^{1,1}\times S^2$, where the sphere could be considered an internal space. Thus, this 
situation can be treated in the same way as in \cite{bmpp-spindel} where a $1+1$ dimensional space time is taken into 
account. There, it is considered the Klein-Gordon equation for a complex scalar field in the presence of an $E$ field, 
but the same reasoning also works for a Dirac field. In this way we obtain a Schr\"odinger equation for a particle in an 
upside down oscillator potential and we can compute the degeneracy factor counting the number of modes whose turning points 
lie within $0<\psi<T$ and $0<\chi<L$, where $T$ and $L$ are the sizes of the space time box over which $E$ is 
nonvanishing. We put $T L = S$, and then we obtain for the degeneracy factor the value of $eE S/2\pi$, 
exactly the same as we will obtain by using the $\zeta$-function approach. We underline that (\ref{wk-ucI}) 
is in strict analogy with the results obtained in \cite{gabriel-spindel} for the scalar case in a 1+1 
Minkowski spacetime, and the terms appearing in (\ref{wk-ucI}) allow an analogous interpretation: the 
first contribution is leading and is a bulk one, proportional to the spacetime volume of the 2D Rindler part 
of the manifold, the latter one is a surface contribution. Cf. also \cite{russo}.


\subsection{The $\zeta$-function approach}

The present case is also analyzed by means of $\zeta$-function techniques, which 
confirm the results obtained in the former approach.\\
For the Dirac operator on the $(\psi,\chi)$-part of the manifold one obtains 
\beq
\slash \!\!\!\! E = \frac{1}{\chi} \tilde{\gamma}_0 \left(\partial_{\psi} - i e E \frac{\chi^2}{2} \right)
+\tilde{\gamma}_1 \left( \partial_{\chi}+ \frac 12 \frac{1}{\chi} \right); 
\eeq
A Liouville unitary transformation 
\beq
(S \Psi) (\psi,\chi) : = \sqrt{\chi} \Psi (\psi,\chi)
\eeq
(i.e. $\Psi (\psi,\chi) = \frac{1}{\sqrt{\chi}} \Phi (\psi,\chi)$), leads to the unitarily related operator 
(still called $\slash \!\!\!\! E$)
\beq
\slash \!\!\!\! E = \frac{1}{\chi} \tilde{\gamma}_0 \left(\partial_{\psi} - i e E \frac{\chi^2}{2} \right)
+\tilde{\gamma}_1 \partial_{\chi},  
\eeq
and then
\beq
E^2 = \frac{1}{\chi^2} \left(\partial_{\psi} - i e E \frac{\chi^2}{2} \right)^2 + \partial^2_{\chi}
+\tilde{\gamma}_0 \tilde{\gamma}_1 \left(i e E\frac{1}{2}+\frac{1}{\chi^2} \partial_{\psi}\right)
\eeq
Substituting $t = \frac{\chi^2}{2}$, performing variable separation 
and keeping into account a unitary ``rotation'' 
transformation which is completely analogous 
to the one performed in the ultracold II case, one obtains for the eigenvalue equation of 
$-E^2$ the following couple of ordinary differential equations:
\beq
t \pa_t^2 \eta_{\pm}(t) + \frac 12 \pa_t \eta_{\pm}(t)
-\frac{1}{4t} \left[ (\omega+\ee t)^2 \pm (\omega - \ee t) - 2 t w^2 \right] \eta_{\pm}(t) = 0. 
\eeq
Let us introduce $g_{\pm} (t)$ through 
\beq
\eta_{\pm}(t) = \exp \left( \pm \frac 12 \ee t \right) t^{\pm \frac 12 \omega} g_{\pm} (t); 
\eeq
then we obtain the following confluent hypergeometric equations:
\beq
t \frac{d^2 g_{\pm}(t)}{dt^2} + \left[\frac 12 \pm (\omega+\ee t) \right] \frac{d g_{\pm}(t)}{dt} +
\frac 12 w^2  g_{\pm}(t) =0.
\eeq
We are looking for solutions $\left( \begin{array}{c} \eta_+ \cr
\eta_- \end{array} \right)$ which belong to the Hilbert space $L^2 [(0,\infty), \frac{dt}{t}]^2$
inherited from the original 4D space (the measure keeps into account the above Liouville 
transformation). 
We need to distinguish two regions for the energy $\omega$.\\
For $\omega <0$, we find
\beq
g(t)_- = \Phi (\frac{-w^2}{2\ee}; \frac 12 -\omega; \ee t ),
\eeq
with the quantization condition
\beq
\frac{-w^2}{2\ee}= -n, 
\eeq
which gives
\beq
w^2_- = 2\ee n.
\eeq
Moreover we find
\beq
g(t)_+ = t^{\frac 12 -\omega}\; \exp (-\ee t)\; \Phi (1-\frac{-w^2}{2\ee}; \frac 32 -\omega; \ee t ),
\eeq
with the quantization condition 
\beq
1-\frac{w^2}{2\ee}= -n, 
\eeq
which gives
\beq
w^2_+ = 2\ee (n+1).
\eeq
Moreover, as in the ultracold II case, we can re-label the eigenvalues in such a way that 
the eigenvalues of $-\slash \!\!\!\! D^2 + \mu^2$ in this region are 
\beq
\lambda^2 = 2 \ee n + \mu_k^2,
\eeq
where again $\mu_k^2 = \mu^2+ 2\Lambda k^2$. 
Also in this case the degeneracy of the $n=0$ case is one half the degeneracy of the 
$n>0$ cases, in full analogy with the ultracold II case. 
For the region $\omega >0$, solutions $\left( \begin{array}{c} \eta_+ \cr
\eta_- \end{array} \right) \in L^2 [(0,\infty), \frac{dt}{t}]^2$ 
correspond to
\beq
g_- (t) = t^{\frac{1}{2}+\omega} \Phi (\frac 12 + \omega- \frac{w^2}{2\ee}; \frac 32 +\omega; \ee t ),
\eeq
and to 
\beq
g_+ (t) = \exp (-\ee t) \Phi (\frac 12 + \omega- \frac{w^2}{2\ee}; \frac 12 +\omega; \ee t ),
\eeq
which are both associated with the quantization condition
\beq
\frac 12 + \omega- \frac{w^2}{2\ee} = -n.
\eeq
The eigenvalues of $-\slash \!\!\!\! D^2 + \mu^2$ are for $\omega>0$ 
\beq
\lambda^2 = (2 n+1) \ee + 2\ee \omega+\mu_k^2.
\eeq
Note that the explicit dependence on $\omega$ appears only for the second region. 
For $\omega<0$, i.e. in the level-crossing region, we have independence on $\omega$ of the integrand. 
The degeneracy factor can be computed as for the ultracold II case giving the same factor. 
As to the term depending on $\omega$, with $\omega>0$, we recall that 
$\sum_{\omega}\mapsto \frac{{\cal T}}{2\pi} \int d\omega$ 
holds. Then we obtain
\beqnl
\frac{1}{2}\zeta_k (s) &=& \frac{\ee S}{2\pi}\left[ (2\ee)^{-s} \zeta_H (\frac{\mu_k^2}{2\ee}, s)  
- \frac 12 \frac{1}{\mu_k^{2 s}} \right]+ 
\frac{{\cal T}}{2\pi} (2\ee)^{-s} 
\frac{1}{s-1}
\zeta_H (\frac 12 (1+\frac{\mu_k^2}{\ee}), s-1).\label{zetaI}
\eeqnl
Note that the first term is the same as for the ultracold II case. Differentiating the second term with 
respect to $s$ in $s=0$ and going back to the Lorentzian signature we obtain the contributions
$$
\frac{{\cal T}}{2\pi} \left[(\log (2ieE)-1) \zeta_H (\frac 12 [1+\frac {\mu_k^2}{ieE}],-1) 
+\frac 12 \log (\mu_k^2)-
\zeta'_H (\frac 12 [1+\frac {\mu_k^2}{ieE}],-1) \right].
$$
To compute the imaginary part of this expression, we first note that
\beq
\zeta_H(a,-1)=-\frac 12 B_2(a)=-\frac 12(a^2- a +\frac 16),
\eeq
($B_2(a)$ stays for the second Bernoulli polynomial) 
so that if $a$ is real
\beq
\imag \zeta_H(\frac 12 +ia,-1)=0.
\eeq
Then, the first term gives the contribution
\beq
\frac 1{8\pi}\left[ \frac 12 \left( \frac {\pi \mu_k^2}{eE} \right)^2 +
\frac {\pi^2}{6} \right] \label{primo}.
\eeq
To compute the second term, let us start from the obvious identity
\beq
\zeta_H(a;z)=\frac 1{2^z} \zeta_H(\frac a2;z)+\frac 1{2^z} \zeta_H(\frac a2+\frac 12;z).
\eeq
Deriving in $z=-1$ and choosing $a=ix$
we get
\beq
\zeta'_H (i\frac x2+\frac 12;-1)=\frac 12 \zeta'_H (ix;-1)-\zeta'_H (i\frac
x2;-1)+\frac 12 \log 2 \zeta_H (ix;-1).
\eeq
Using $\imag \zeta_H (ix;-1)=\frac x2$ and formula (20) in \cite{zerbin} we get
\begin{align}\label{formola}
\imag \zeta'_H (i\frac x2+\frac 12;-1)= & \frac 1{8\pi} \left[ -\frac {\pi^2}6 -\frac
12 (x\pi)^2 -\sum_{n=1}^\infty \frac {e^{-2n\pi x}}{n^2}
+2 \sum_{n=1}^\infty \frac {e^{-n\pi x}}{n^2} \right] \cr
& =\frac 1{8\pi} \left[ -\frac {\pi^2}6 -\frac 12 (x\pi)^2 -2 \sum_{n=1}^\infty
(-1)^n \frac {e^{-n\pi x}}{n^2} \right].
\end{align}
Using the expression
\beq
\sum_{n=1}^\infty (-1)^n \frac {e^{-n\pi x}}{n^2}={\mathrm{Li}}_2 (-e^{-\pi x}),
\eeq
and adding (\ref{primo}) and the derivative of the first term in (\ref{zetaI}), we finally get
\beq
\frac{1}{2}\imag \zeta'_k (0) = \frac{e E S}{2\pi} \left[
-\frac{1}{2} \log \left(1-\exp \left(-\frac{\pi \mu_k^2}{\ee}\right)\right)\right] - \frac{{\cal T}}{2\pi} 
\frac{1}{4\pi} {\mathrm{Li}}_2 \left(-\exp \left(-\frac{\pi \mu_k^2}{\ee}\right)\right).
\eeq
This result coincides with (\ref{wk-ucI}). It is remarkable that the transmission coefficient 
approach involves only an integral over the level-crossing region, whereas the $\zeta$-function 
calculation requires a control over the whole spectrum, and is quite more tricky, but it provides 
also more information, being a priori the complete 1-loop effective action made available by the 
latter approach.

\subsection{Instability of the thermal state}
\label{inst-ucI}

According to the general discussion of section \ref{vac-inst}, we get
\beq
<N_l^{out}>_{\beta_h} = \frac{1}{1+\exp [2\pi (\omega-\varphi^+)]} 
+  
\frac {e^{-\frac {\pi \mu_k^2}{2eE}} \cosh (\pi \omega)}{\cosh
[\pi(\omega-\frac {\mu_k^2}{2eE})]} \frac 12 \left( 
\tanh [\pi (\omega-\varphi^+)]+\tanh [\pi (|\omega|+\varphi^-)] \right), 
\eeq
where the latter term is the net effect associated 
with the pair production induced by the presence of an electrostatic field. 
Notice however that the potential $\varphi^+=\varphi^-$ is ill-defined, being infinite 
unless a spatial cut-off is introduced at $\chi=\chi_0>0$.

\section{Nariai case}
\label{nariai}

We now consider the more general case, that is the electrically charged Nariai solution.
The manifold is described by the metric \cite{romans,bousso,mann}
\beq
ds^2 = \frac{1}{A} (-\sin^2 (\chi) d \psi^2 + d\chi^2) + \frac{1}{B} (d\theta^2+
\sin^2 (\theta) d\phi^2)
\label{nariai-metric}
\eeq
with $\psi \in \RR, \chi\in (0,\pi)$, and the constants 
$B=\frac{1}{2 Q^2}\left( 1-\sqrt{1-12 \frac{Q^2}{L^2}} \right)$, $A=\frac{6}{L^2}-B$
are such that $\frac{A}{B}<1$, and $L^2:= \frac{3}{\Lambda}$. The black hole horizon 
occurs at $\chi=\pi$. This manifold differs 
from the ultracold cases because it has finite spatial section. 
In the Euclidean version, 
it corresponds to two spheres characterized by different radii. 
One finds the following non-vanishing Christoffel symbols $\Gamma_{01}^0 = \cot (\chi),
\Gamma_{00}^1 = \sin (\chi) \cos (\chi), \Gamma_{33}^2 = -\sin (\theta) \cos (\theta),
\Gamma_{23}^3 = \cot (\theta)$. 
For the gauge potential we can choose 
$A_{i} = -Q \frac{B}{A} \cos (\chi) \delta_{i}^0$.\\
As the situation is more difficult, in this section we will provide a more 
detailed exposition.

\subsection{The transmission coefficient approach}

By variable separation, as performed in \cite{belcaccia-rnds}, we obtain the 
following reduced Hamiltonian (where $\mu,e$ are the fermion mass and charge, as before):
\beq
h_k = \left[
\begin{array}{cc}
e Q  \frac{B}{A} \cos (\chi) - \frac{\mu}{\sqrt{A}} \sin (\chi)
& \sin(\chi) \pa_{\chi} + \sqrt{\frac{B}{A}} \sin (\chi) k \cr
-\sin(\chi) \pa_{\chi} + \sqrt{\frac{B}{A}} \sin (\chi) k
& e Q  \frac{B}{A} \cos (\chi) + \frac{\mu}{\sqrt{A}} \sin (\chi)
\end{array}
\right].
\label{hk-nariai}
\eeq
We introduce, with the aim of simplifying the notation, the following 
definition:
\beq
E:= Q  \frac{B}{A},
\eeq
which corresponds to $\frac{1}{A}$ times the maximum value for the intensity of the
electrostatic field. Moreover, we assume for definiteness $\ee>0$, and also 
we adopt the following redefinitions:
\beqnl
\frac{1}{\sqrt{A}} \mu &\mapsto& \mu,\cr
\sqrt{\frac{B}{A}} k &\mapsto& k.
\label{redef}
\eeqnl
Then the Dirac equation in the Hamiltonian form is 
\beqna
\left(\begin{array}{cc}
eE\cos \chi -\mu \sin \chi & \sin \chi \partial_\chi +k\sin \chi \\
-\sin \chi \partial_\chi +k\sin \chi & eE\cos \chi +\mu \sin \chi \end{array} \right)
\left( \begin{array}{c} \psi_1 \\ \psi_2 \end{array} \right)=\omega \left(
\begin{array}{c} \psi_1 \\ \psi_2 \end{array} \right).
\eeqna
Using the coordinate $t=-\cos \chi$ we can write it in the form
\beq
\left[(eEt+\omega) \mathbb{I}_2 +\mu \sqrt {1-t^2} \sigma_3 -i(1-t^2)\sigma_2
\partial_t -k\sqrt{1-t^2} \sigma_1 \right]  \psi= 0,
\eeq
where $\Psi=\left(^{\psi_1}_{\psi_2}\right)$. 
Let us take the
unitary transformation $\Psi=e^{-i\frac \pi4 \sigma_1} \xi$;  
then
\begin{eqnarray}
&& [(1-t^2)\partial_t +i (eEt+\omega)]\xi_1-(\mu+ik) \sqrt{1-t^2} \xi_2=0
\label{xi1},\\
&& [(1-t^2)\partial_t -i (eEt+\omega)]\xi_2-(\mu-ik) \sqrt{1-t^2} \xi_1=0. \label{xi2}
\end{eqnarray}
{F}rom (\ref{xi1}) we find
\beq
(\mu+ik)\xi_2 =\sqrt{1-t^2} \xi_1'+i\ \frac {eEt+\omega}{\sqrt{1-t^2}} \xi_1,
\label{seconda}
\eeq
which inserted into (\ref{xi2}) gives
\beq
(1-t^2) \xi_1''-t\xi_1' -(\mu^2 +k^2 -ieE)\xi_1 +\frac 1{1-t^2}
[(eEt+\omega)^2+i(eEt+\omega)t]\xi_1=0.\label{star}
\eeq
Looking at the behavior of this equation at the singular points $t=\pm 1$ we see that it is 
convenient to define a function $\zeta$ such that
\begin{eqnarray}
&& \xi_1(t)=(1-t)^\alpha (1+t)^\beta \zeta((1+t)/2), \\
&& \alpha \in \{ \frac 12 (1-ieE -i\omega), \frac i2 (eE+\omega) \}, \qquad\ \beta
\in \{ \frac 12 (1-ieE +i\omega), \frac i2 (eE-\omega) \}.
\end{eqnarray}
Setting $z=(1+t)/2$ and choosing $\alpha=\frac i2 (eE+\omega)$ and $\beta=\frac i2 (eE-\omega)$, we find
\beq
z(1-z)\zeta'' +[2\beta+\frac 12 -z(1+2i eE)] \zeta' -[\mu^2+k^2] \zeta=0,
\eeq
which is an hypergeometric differential equation. 
Note that the solution regular in $z=0$ of this equation provides a solution of
(\ref{star}) regular in $t=-1$. Moreover, (\ref{star}) is invariant under the combination of 
complex conjugation with the transformation $(t,E)\rightarrow (-t,-E)$. 
Thus, we are able to get the solution regular in $t=1$ applying this 
transformation to the solution regular in $t=-1$.
This gives
\begin{align}
& \xi_1(t)=c_1 \left( \frac {1-t}2 \right)^{i\frac {eE+\omega}2} \left( \frac
{1+t}2 \right)^{i\frac {eE-\omega}2}
{}_2 F_1(ieE +i\sqrt \Delta , ieE -i\sqrt \Delta ; \frac 12 +i(eE-\omega); \frac {1+t}2) \cr
& \qquad\ +c_2 \left( \frac {1-t}2 \right)^{i\frac {eE+\omega}2} \left( \frac
{1+t}2 \right)^{i\frac {eE-\omega}2}
{}_2 F_1(ieE +i\sqrt \Delta , ieE -i\sqrt \Delta ; \frac 12 +i(eE+\omega); \frac {1-t}2),
\end{align}
where $\Delta:= \mu^2+k^2+e^2E^2$, and ${}_2 F_1$ is the well-known Gauss hypergeometric function.\\
Using (\ref{seconda}) and the relations
\begin{align*}
& -\sqrt{1-t^2} \frac d{dt}\left[ \left( \frac {1-t}2 \right)^{i\frac {eE+\omega}2}
\left( \frac {1+t}2 \right)^{i\frac {eE-\omega}2} \right]
-i\ \frac {eEt+\omega}{\sqrt{1-t^2}} \left[ \left( \frac {1-t}2 \right)^{i\frac
{eE+\omega}2} \left( \frac {1+t}2 \right)^{i\frac {eE-\omega}2}
\right]=0 \\
& {}_2 F_1'(a,b;c;z)=\frac {ab}c {}_2 F_1(a+1,b+1;c+1;z),
\end{align*}
we get
\begin{align}
\xi_2(t)=&c_1 \frac {2(\mu-ik)}{1+2i(eE-\omega)} \left( \frac {1-t}2
\right)^{i\frac {eE+\omega}2+\frac 12}
\left( \frac {1+t}2 \right)^{i\frac {eE-\omega}2+\frac 12} \\ \nonumber
&{}_2 F_1(ieE +i\sqrt \Delta +1, ieE -i\sqrt \Delta +1; \frac 32 +i(eE-\omega); \frac
{1+t}2) \\ \nonumber
& -c_2 \frac {2(\mu-ik)}{1+2i(eE+\omega)} \left( \frac {1-t}2
\right)^{i\frac {eE+\omega}2 +\frac 12}
\left( \frac {1+t}2 \right)^{i\frac {eE-\omega}2+\frac 12}\\ \nonumber
&{}_2 F_1(ieE +i\sqrt \Delta +1, ieE -i\sqrt \Delta+1 ; \frac 32 +i(eE+\omega); \frac {1-t}2).
\end{align}
As in the previous cases, we can look at the asymptotic behaviors at infinities, in
the coordinate $x=\log \tan \frac \chi2$. For $x\rightarrow -\infty$
we get
\begin{align}
& \xi_1(x) \approx  \left[ c_1+c_2 \frac {\Gamma(\frac 12 +i(eE+\omega))
\Gamma(\frac 12 -i(eE-\omega))}{\Gamma(\frac 12 +i\omega -i\sqrt \Delta)\Gamma(\frac
12 +i\omega +i\sqrt \Delta)}\right] e^{i(eE-\omega)x}
+e^{-i(eE-\omega)x} O(e^{x}),\\
& \xi_2(x) \approx -c_2\ \frac {2(\mu-ik)}{1+2i(eE+\omega)} \frac {\Gamma(\frac 32
+i(eE+\omega))
\Gamma(\frac 12 +i(eE-\omega))}{\Gamma(1 +ieE -i\sqrt \Delta)\Gamma(1 +ieE +i\sqrt
\Delta)} e^{-i(eE-\omega)x}
+e^{i(eE-\omega)x} O(e^{x}).
\end{align}
For $x\rightarrow +\infty$:
\begin{align}
& \xi_1(x) \approx  \left[ c_2+c_1 \frac {\Gamma(\frac 12 +i(eE-\omega))
\Gamma(\frac 12 -i(eE+\omega))}{\Gamma(\frac 12 -i\omega -i\sqrt \Delta)\Gamma(\frac
12 -i\omega +i\sqrt \Delta)}\right]e^{-i(eE+\omega)x}
+e^{i(eE+\omega)x} O(e^{-x}),\\
& \xi_2(x) \approx c_1\ \frac {2(\mu-ik)}{1+2i(eE-\omega)} \frac {\Gamma(\frac 32
+i(eE-\omega))
\Gamma(\frac 12 +i(eE+\omega))}{\Gamma(1 +ieE -i\sqrt \Delta)\Gamma(1 +ieE +i\sqrt
\Delta)} e^{i(eE+\omega)x}
+e^{-i(eE+\omega)x} O(e^{-x}).
\end{align}
To compute the transmission and reflection coefficient let us write the asymptotic 
expressions for $\xi$ as
\begin{eqnarray}
&& \xi_1(x)^- \approx  \left[ c_1+c_2 \alpha_0 \right] e^{i(eE-\omega)x}
+e^{-i(eE-\omega)x} O(e^{x}),\\
&& \xi_2(x)^- \approx -c_2\ \beta_0 e^{-i(eE-\omega)x}
+e^{i(eE-\omega)x} O(e^{x}),
\end{eqnarray}
for $x\rightarrow -\infty$, and for $x\rightarrow \infty$
\begin{eqnarray}
&& \xi_1(x)^+ \approx  \left[ c_2+c_1 \alpha_1 \right]e^{-i(eE+\omega)x}
+e^{i(eE+\omega)x} O(e^{-x}),\\
&& \xi_2(x)^+ \approx c_1\ \beta_1 e^{i(eE+\omega)x}
+e^{-i(eE+\omega)x} O(e^{-x}).
\end{eqnarray}
By taking into account correctly group velocity and imposing that there is only 
incoming wave at $x=-\infty$, one finds
\beqnl
|T_k (\omega)|^2&=& 
\frac{\cosh [\pi(eE-\omega)]\cosh [\pi(eE+\omega)]}{\cosh [\pi(\sqrt
\Delta-\omega)]\cosh [\pi(\sqrt \Delta+\omega)]}\cr
&=&\frac{\cosh [2\pi eE]+\cosh [2 \pi \omega]}{\cosh [2 \pi \sqrt \Delta]+\cosh [2
\pi \omega]},
\eeqnl
which has the expected property to satisfy $|T|^2<1$ and gives the mean number of 
created pairs for unit time and unit volume. Moreover, we get 
\beq
|R_k (\omega)|^2=\frac {\sinh[\pi(\sqrt \Delta -eE)]\sinh[\pi(\sqrt \Delta +eE)]}{\cosh
[\pi(\sqrt \Delta-\omega)]\cosh [\pi(\sqrt \Delta+\omega)]},
\eeq
and the property $|T_k (\omega)|^2+|R_k (\omega)|^2 = 1$ is verified.\\
We recall that one has to reinstate in the above formulas for $|T|^2$ the original values 
for $\mu$ and $k$ (cf. eqn. (\ref{redef})). Then $\Delta = \frac{\mu^2}{A}+ \frac{B}{A} k^2 + (e E)^2$.\\ 

Comparing with \cite{belcaccia-rnds}, the limit as $eE\to \infty$ leads 
to the WKB approximation (and to the limit $|T_k (\omega)|^2\to 1^-$). This is the actual approximation 
where the WKB approximation works well. We can notice that the above limit 
actually means $Q \frac{B}{A} \to \infty$. Being
\beq
\frac{B}{A} = \frac{1}{\sqrt{1-4 \Lambda Q^2}},
\eeq
one can obtain the above limit as $Q^2 \to \left(\frac{1}{4\Lambda}\right)^-$.\\

In order to determine the imaginary part of the effective action, we refer again to (\ref{im-wk}).  
We do not perform the sum over
$k$, and then we calculate 
\beq
W_k = -\frac 12 \sum_{\omega} \log (1-|T_k (\omega)|^2).
\eeq
We sum only over the level-crossing region, because only there particle creation
is expected to be present, and then only there an instability for the vacuum should occur.
This region is $-eE \leq \omega \leq eE$ so that we have to perform the following integral:
\beq
I:=\int_{-eE}^{eE} d\omega \log (1 -
\frac{\cosh [2\pi eE]+\cosh [2 \pi \omega]}{\cosh [2 \pi \sqrt \Delta]+\cosh [2 \pi
\omega]}),
\eeq
where the dependence on $k$ is implicit in $\Delta$; the integral can be rewritten
as follows:
\beq
I = 2 eE \log (\cosh [2\pi \sqrt \Delta]-\cosh [2 \pi eE])-II,
\eeq
where
\beq
II:=\int_{-eE}^{eE} d\omega \log (\cosh [2\pi \sqrt \Delta]+\cosh [2 \pi \omega]) =
\frac{1}{2\pi} \int_{-2\pi eE}^{2\pi eE} dy \log (p+\cosh [y]),
\eeq
with $p:=\cosh [2\pi \sqrt \Delta]$. By taking into account that
$p+\cosh [y] = \frac{1}{2} (2p + \exp(y) +\exp(-y))=\frac{1}{2} \exp(y) ( \exp(-
2y)+2 p \exp(- y)+1) =
\frac{1}{2} \exp(y) (\exp(-y) +p -\sqrt{p^2-1}) (\exp(-y) +p +\sqrt{p^2-1})$ and that
$p -\sqrt{p^2-1}=\exp (-2 \pi \sqrt{\Delta})$ and $p +\sqrt{p^2-1}=\exp (2 \pi
\sqrt{\Delta})$,
one finds
\beq
II = \frac{1}{2\pi} \int_{-2\pi eE}^{2\pi eE} dy \left[y-\log 2 +
\log(\exp(-y)\exp(-2 \pi \sqrt{\Delta}) +1)
+\log(\exp(-y)\exp(2 \pi \sqrt{\Delta}) +1) \right].
\eeq
The following result is useful:
\begin{align}
\int dy\log (\exp(-y+\delta)+1) = &\frac{1}{2} y^2 + y \log(\exp(-y+\delta)+1) -y 
\log (\exp(y-\delta)+1) \cr
&- {\mathrm{Li}}_2 
(-\exp (y-\delta))\cr 
= &-\frac{1}{2} y^2 + y \delta-
{\mathrm{Li}}_2 (-\exp (y-\delta)).
\end{align}
As a consequence, with simple manipulations, we get
\begin{align}
II &= \frac{1}{2\pi} \left[-2 (2\pi eE)\log 2 + {\mathrm{Li}}_2 (-\exp [-2\pi
(\sqrt{\Delta}+eE)])
- {\mathrm{Li}}_2 (-\exp [2\pi (\sqrt{\Delta}+eE)]) \right. \cr
&+\left.  {\mathrm{Li}}_2 (-\exp [2\pi (\sqrt{\Delta}-eE)])- {\mathrm{Li}}_2 (-\exp
[-2\pi (\sqrt{\Delta}-eE)]) \right],
\end{align}
and then, by taking into account that $\sum_{\omega}\mapsto\frac{{\cal T}}{2\pi} \int d\omega$, we get 
\begin{align}
W_k &= - \frac{{\cal T}}{2\pi} eE \log (2\cosh [2\pi \sqrt{\Delta}]-2\cosh [2 \pi eE]) -
\frac{{\cal T}}{2\pi} 
\frac{1}{4\pi} \left[- {\mathrm{Li}}_2 (-\exp [-2\pi (\sqrt{\Delta}+eE)])\right. \\ \nonumber
&+ \left. {\mathrm{Li}}_2 (-\exp [2\pi (\sqrt{\Delta}+eE)])
- {\mathrm{Li}}_2 (-\exp [2\pi (\sqrt{\Delta}-eE)])+ {\mathrm{Li}}_2 (-\exp [-2\pi
(\sqrt{\Delta}-eE)]) \right].
\end{align}
In the Nariai case, it is not so straightforward to distinguish between bulk and surface parts 
of the given $W_k$. By keeping into account that $E\propto \frac{1}{A}$ and that the volume 
of the (Euclidean) 2D $(\psi,\chi)$-part of the metric is $\frac{4\pi}{A}$, one realizes that the 
first above contribution is leading and is a bulk one and the latter is again a surface contribution.

\subsection{The $\zeta-$function approach}

This case requires more efforts in order to be tackled with $\zeta-$function techniques, 
in particular it needs techniques which have been developed only recently \cite{caccia-zeta}, 
and we sketch herein a more heuristic approach which still leads to the desired results.\\

As to the operator $\slash \!\!\!\! E$ on the $(\psi,\chi)$-part of the manifold, one obtains 
\beq
\slash \!\!\!\! E = 
\frac{\sqrt{A}}{\sin \chi} \tilde{\gamma}_0 \left(\partial_{\psi} + i e E \cos \chi \right)
+\sqrt{A} \tilde{\gamma}_1 \left(\partial_{\chi} +\frac 12 \cot \chi \right).
\eeq
After a Liouville unitary transformation $\Psi (\psi,\chi) = \frac{1}{\sqrt{\sin \chi}} \phi (\psi,\chi)$, 
we get the simplified form (again called $\slash \!\!\!\! E$) 
\beq
\slash \!\!\!\! E = 
\frac{\sqrt{A}}{\sin \chi} \tilde{\gamma}_0 \left(\partial_{\psi} + i e E \cos \chi \right)
+\sqrt{A} \tilde{\gamma}_1 \partial_{\chi}; 
\eeq
moreover, 
\beq
E^2 = \frac{A}{\sin^2 \chi} \left(\partial_{\psi} + i e E \cos \chi \right)^2 + A \partial^2_{\chi} + 
A \tilde{\gamma}_0 \tilde{\gamma}_1 \left( \frac{\cos \chi}{\sin^2 \chi }
\left(\partial_{\psi} + i e E \cos \chi \right) + i \ee \right).
\eeq
Substituting $t = -\cos \chi$ one obtains 
\beq
E^2 = \frac{A}{1-t^2} \left(\partial_{\psi} - i e E t \right)^2 + 
A (1-t^2) \pa_t^2 - A t \pa_t +  
A \tilde{\gamma}_0 \tilde{\gamma}_1 \left( -\frac{t}{1-t^2 }
\left(\partial_{\psi} - i e E t \right) + i \ee \right).
\eeq
After variable separation, 
a reduction and a rotation as in the previous cases, we obtain for the eigenvalue problem 
of $-E^2$ the following couple of differential equations:
\beq
(1-t^2) \pa_t^2 \eta_{\pm} -  t \pa_t \eta_{\pm} \pm \ee \eta_{\pm} + 
\frac{1}{1-t^2} \left[-(\omega + \ee t)^2 \pm t (\omega + \ee t) \right]\eta_{\pm}+ \frac{w^2}{A}\eta_{\pm}=0.
\eeq
We choose
\beqnl
\eta_+ (t) &=& (1-t)^{\frac{\ee + \omega}{2}} (1+t)^{\frac{\ee - \omega}{2}} g_+ (t),\\
\eta_- (t) &=& (1-t)^{\frac{-\ee - \omega}{2}} (1+t)^{\frac{-\ee + \omega}{2}} g_- (t).
\eeqnl
Then by choosing $t = 2 z -1$ 
we obtain the following couple of hypergeometric equations:
\beq
z (1-z) \frac{d^2 g_+ (z)}{dz^2} + \left(\ee -\omega+\frac 12 - (2\ee +1) z\right) \frac{d g_+ (z)}{dz}
+\frac{w^2}{A} g_+ (z) =0,
\label{hyperp}
\eeq
with $a_+ = \ee+\sqrt{\frac{w^2}{A}+(\ee)^2}, b_+ =\ee-\sqrt{\frac{w^2}{A}+(\ee)^2}, 
c_+=\ee-\omega+\frac 12$,  
and 
\beq
z (1-z) \frac{d^2 g_- (z)}{dz^2} + \left(-\ee +\omega+\frac 12 + (2\ee -1) z\right) \frac{d g_- (z)}{dz}
+\frac{w^2}{A} g_- (z) =0, 
\label{hyperm}
\eeq
with $a_- = -\ee+\sqrt{\frac{w^2}{A}+(\ee)^2}, b_- =-\ee-\sqrt{\frac{w^2}{A}+(\ee)^2}, 
c_-=-\ee+\omega+\frac 12$. 
We are looking for solutions $\left( \begin{array}{c} \eta_+ \cr
\eta_- \end{array} \right) \in L^2 [(0,1), \frac{dz}{z (1-z)}]^2$.  
It is not difficult to realize that this condition depends on $\omega$, and
three regions can be identified.\\ 
One can show that 
\beq
g_+ (z) ={}_2F_1 (a_+,b_+;c_+;z)
\eeq
with the quantization condition  
\beq
b_+ = -n,
\eeq
which leads to 
\beq
w^2_+ = A (\ee + n )^2 - A (\ee)^2,
\eeq
together with 
\beq
g_- (z) = z^{1-c_-} (1-z)^{c_- - (a_- + b_-)} 
{}_2F_1 (1-a_-, 1-b_-; 2-c_-; z)
\eeq
with the quantization condition 
\beq
1-a_- = -n, 
\eeq
which leads to 
\beq
w^2_- = A (\ee + n + 1 )^2 - A (\ee)^2,
\eeq
correspond to solutions of (\ref{hyperp}), (\ref{hyperm}) respectively which allow to 
get $\left( \begin{array}{c} \eta_+ \cr
\eta_- \end{array} \right) \in L^2 [(0,1), \frac{dz}{z (1-z)}]^2$ in the 
interval $-\ee < \omega < \ee$. We can re-label the eigenvalue and obtain for $-D^2+\mu^2$ 
\beq
\lambda^2 = A (\ee + n)^2 - A (\ee)^2 + \mu^2 + B k^2.
\eeq
Also in this case, one has to take into account that degeneracy for $n=0$ is one half 
the degeneracy for $n>0$.\\ 
In the region $\omega> \ee$ one can show that the couple 
\beq
g_+ (z) = z^{1-c_+} {}_2F_1 (a_+ - c_+ +1, b_+ - c_+ +1; 2-c_+; z)
\eeq
and 
\beq
g_- (z) = (1-z)^{c_- - (a_- + b_-)} {}_2F_1 (c_- - a_-, c_- - b_-; c_-;z)
\eeq
under the quantization conditions $b_+ - c_+ +1 = -n$ and $c_- - a_- = -n$, which both lead to 
\beq
w^2 = A \left(\omega+\frac 12 +n \right)^2 - A (\ee)^2    
\eeq
provides a
solution which is in $L^2 [(0,1), \frac{dz}{z (1-z)}]^2$.
Then the eigenvalue of $-D^2+\mu^2$ becomes
\beq
\lambda^2 = A (\omega + n + \frac 12)^2 - A (\ee)^2+\mu^2 + B k^2.
\eeq
There is still the region $\omega < -\ee$ to be explored, where 
\beq
g_+ (z) = (1-z)^{c_+ - (a_+ + b_+)} {}_2F_1 (c_+ -b_+ ,c_+ -a_+; c_+ - (a_+ + b_+) + 1; 1-z),  
\eeq
and 
\beq
g_- (z) = z^{1-c_-} {}_2F_1 (1 + b_- - c_- , 1 + a_- - c_- ; a_- + b_- + 1 - c_-; 1-z)   
\eeq
with the quantization conditions $c_+ - a_+ = -n$ and $1 + b_- - c_- = -n$, 
which both correspond to 
\beq
w^2 = A \left( -\omega+\frac 12 +n \right)^2 - A (\ee)^2,
\eeq
satisfy the property to be eigenfunctions of the operator $-E^2$. 
The eigenvalues of $-D^2+\mu^2$ are 
\beq
\lambda^2 = A (-\omega + n + \frac 12)^2 - A (\ee)^2 +\mu^2 + B k^2. 
\eeq
Note that, for $|\omega| > \ee$ the eigenvalues of $-D^2+\mu^2$
can be written as follows:
\beq
\lambda^2 = A (|\omega| + n + \frac 12)^2 +\mu^2 + B k^2- A (\ee)^2,
\eeq
and then the integration for $|\omega| > \ee$ is symmetric.\\
For the heat kernel we get
\beq
K (s) = \sum_k g(k) K_k (s),
\eeq
with
\beqnl
K_k (s) &=&2\frac{{\cal T}}{2\pi}\left\{  2 \int_{\ee}^{\infty } d\omega \sum_{n=0}^{\infty}
\exp \left[ -A \left( (\omega + \frac 12 +n)^2 + \frac{\mu_k^2}{A} -(\ee)^2 \right)
s \right]\right.\cr 
&+& \left. 2\ee \sum_{n=0}^{\infty}
\exp \left[ -A \left( (\ee +n)^2 + \frac{\mu_k^2}{A} -(\ee)^2 \right) s \right] 
-\ee \exp \left( -\mu_k^2 s \right)
\right\},
\eeqnl
where $\mu^2_k = \mu^2 + B k^2$. Correspondingly, we obtain
\beqnl
\frac{1}{2}\zeta_k (s) &=&\frac{{\cal T}}{2\pi}\left[
2\int_{\ee}^\infty\sum_n \frac{d\omega}{A^s\left[\left(n+\frac{1}{2}+\omega\right)^2+
\frac{\mu_k^2}{A}-(\ee)^2 \right]^s} +\right.\cr
&+& \left. \left(2 \ee \right) \left( \sum_n
\frac{1}{A^s\left[\left(n+\ee\right)^2+\frac{\mu_k^2}{A}-(\ee)^2
\right]^s} - \frac 12 \frac{1}{\mu_k^{2s}}\right) \right].
\eeqnl
It is convenient to introduce the functions
\beq
\sigma_k(s;z):=\sum_n \frac{1}{A^s\left[\left(n+\frac{1}{2}+\ee +z\right)^2+\frac{\mu_k^2}{A}
-(\ee)^2\right]^s},
\eeq
so that
\beq\label{sigmak}
\frac{1}{2}\zeta_k(s)= \frac{{\cal T}}{2\pi}\left[2\ee \left(\sigma_k(s;-\frac 12)- \frac 12 \frac{1}{\mu_k^{2s}}\right)
+2\int_0^\infty \sigma_k(s;z) dz\right].
\eeq
It is also useful to define $\alpha = \frac 12 +\ee$ and $\beta^2 =
\frac{\mu_k^2}{A} -(\ee)^2$. 
Using the Abel-Plana formula\footnote{Abel-Plana formula is:
$\sum_{n=0}^{\infty}f(n)=\frac{1}{2}f(0)+\int_0^\infty f(x)dx+i\int_0^\infty
\frac{f(ix)-f(-ix)}{e^{2\pi x}-1}$.} we get
\begin{align} \label{plana1}
\sigma_k(s;z) =
&\frac{1}{2}  \frac{1}{A^s\left[\left(\alpha+z\right)^2+\beta^2 \right]^s} +
  \int_0^\infty
\frac{dx}{A^s\left[\left(x+\alpha+z\right)^2+\beta^2\right]^s} \\ \nonumber
&+ i \int_0^\infty dx\left\{
\frac{1}{A^s\left[\left(ix+\alpha+z\right)^2+\beta^2\right]^s}\right. 
\left.
 -\frac{1}{A^s\left[\left(-ix+\alpha+z\right)^2+\beta^2\right]^s}
\right\}
\frac{1}{e^{2\pi x}-1}.
\end{align}
To compute the effective action, we need to compute the derivative of
$\sigma_k(s;z)$ with respect to $s$, in $s=0$.
For the last term of \eqref{plana1} we get
\beq \label{der1}
- i \int_0^\infty dx\,\{\ln [(ix+\alpha+z)^2+\beta^2]-\ln
[(-ix+\alpha+z)^2+\beta^2]\}\frac{1}{e^{2\pi x}-1}.
\eeq
Note that this coincides with
\begin{align}
 i\frac{d}{ds}|_{s=0}\int_0^\infty dx & \frac 1{A^{\gamma s}} \left\{
\frac{1}{(ix+\alpha+z+i\beta)^s}  - \frac{1}{(-ix+\alpha+z+i\beta)^s}
\right. \cr
& \left. + \frac{1}{(ix+\alpha+z-i\beta)^s} -\frac{1}{(-ix+\alpha+z-i\beta)^s}
\right\}\frac{1}{e^{2\pi x}-1},
\end{align}
where $\gamma$ is an arbitrary constant. The result does not depend on $\gamma$ and
we could choose it equal to $0$. However, it is convenient to choose $\gamma=\frac 12$.
Applying Plana's formula to this integrals we see that
\beq
\frac{\partial}{\partial s}|_{s=0} \sigma_k(s;z)=\frac{\partial}{\partial s}|_{s=0}
\tilde \sigma_k(s;z),
\eeq
where
\begin{align} \label{plana2}
\tilde \sigma_k(s;z) =
&\frac{1}{2}  \frac{1}{A^s\left[\left(\alpha +z\right)^2+\beta^2\right]^s} +
  \int_0^\infty \frac{dx}{A^s\left[\left(x+\alpha+z\right)^2+\beta^2 \right]^s}
\cr
& + \frac 1{A^{s/2}} \sum_{n=0}^\infty \left \{\frac{1}{(n+\alpha+z+i\beta)^s}+
\frac{1}{(n+\alpha+z-i\beta)^s} \right\} \cr
&-\frac{1}{2}  \frac 1{A^{s/2}} \left\{ \frac{1}{(\alpha+z+i\beta)^s}+
\frac{1}{(\alpha+z-i\beta)^s}  \right\} \cr
&-  \frac 1{A^{s/2}} \int_0^\infty dx\left \{\frac{1}{(x+\alpha+z+i\beta)^s}+
\frac{1}{(x+\alpha+z-i\beta)^s}  \right\}.
\end{align}
We now note that, thanks to our choice for $\gamma$, collecting the second and the
last terms under the integral and deriving with respect
to $s$ in $s=0$ we obtain a vanishing term (see Lemma 1 in \cite{caccia-zeta}).
The same happens for the first and the fourth term so that only the third term
contribute to the derivative. Our conclusion is that
\beq
\frac{\partial}{\partial s}|_{s=0} \sigma_k(s;z)=\frac{\partial}{\partial s}|_{s=0}
\left[  \frac 1{A^{s/2}}
(\zeta_H (\alpha+z+i\beta,s)+\zeta_H (\alpha+z-i\beta,s)) \right].
\eeq
Using (\ref{sigmak}) we conclude that
\beq
\frac{\partial}{\partial s}|_{s=0} \zeta_k(s)=\frac{\partial}{\partial s}|_{s=0}
\hat \zeta_k(s),
\eeq
where
\beq
\frac{1}{2}\hat \zeta_k(s):= \frac{{\cal T}}{2\pi}\left[2\ee\left( \hat \sigma_k(s;-\frac 12)
- \frac 12 \frac{1}{\mu_k^{2s}}\right)
+2\int_0^\infty \hat \sigma_k(s;z) dz \right],
\eeq
with
\beq
\hat \sigma_k(s;z)=  \frac 1{A^{s/2}} (\zeta_H (\alpha+z+i\beta,s)+\zeta_H (\alpha+z-i\beta,s)).
\eeq
Thus
\beqnl
\frac{1}{2}\hat{\zeta}_k (s) &=& \frac{{\cal T}}{2\pi}\left\{ 2\ee  \frac{1}{A^{\frac{s}{2}}} \left[
\zeta_H (\alpha -\frac 12 +i \beta, s) 
+ \zeta_H (\alpha -\frac 12 -i \beta, s)
\right]-\ee \frac{1}{\mu_k^{2s}}\right.\cr
&+&\left. 2 \frac{1}{A^{\frac{s}{2}}} \frac{1}{s-1} \left[
\zeta_H (\alpha +i \beta, s-1) + \zeta_H (\alpha -i \beta, s-1) \right]\right\}.
\eeqnl
Then we get (we momentarily omit the index $k$ from some formulas below)
\beqnl
\frac{1}{2}\zeta' (0) &=&\frac{{\cal T}}{2\pi}\left\{ 2(\ee)^2 \log A -\ee \log A  + 2\ee \log
\frac{\Gamma (\ee+ i \beta) \Gamma (\ee- i \beta)}{2\pi} \right.\cr
&+& \left. \ee \log (\mu_k^2) + (2+\log A) \left[
\zeta_H (\alpha +i \beta, -1) + \zeta_H (\alpha -i \beta, -1) \right] \right.\cr
&-& 2 \left. \left[
\zeta'_H (\alpha +i \beta, -1) + \zeta'_H (\alpha -i \beta, -1) \right]\vphantom{\frac 12}\right\},
\eeqnl
where we used the well known relations
\beqnl
\zeta_H(a,0)=\frac 12 -a, \qquad\ \zeta'_H(a,0)=\log \frac {\Gamma(a)}{\sqrt {2\pi}}. 
\eeqnl
By going back to Lorentzian signature, through $\ee \mapsto i \ee$, we get
\beqnl
\frac 12 \imag  \zeta' (0)  &=& \frac{{\cal T}}{2\pi}\left[-\ee \log A - \ee \log (\Delta - (\ee)^2 )
-\ee \log \left( 2 \cosh[2 \pi \sqrt{\Delta}] -2 \cosh [2\pi \ee] \right) \right.\cr
&+& \left. \ee \log (\mu_k^2)- 2 \imag \left[
\zeta'_H (\alpha +i \beta, -1) + \zeta'_H (\alpha -i \beta, -1) \right] \vphantom{\frac 12}\right].
\eeqnl
To compute the last two terms we can start from (\ref{formola}) and use the identity
\beqnl
-\frac 12 \log^2 (-z)-\frac {\pi^2}6={\mathrm{Li}}_2 (z)+{\mathrm{Li}}_2(1/z),
\eeqnl
to obtain the relation
\beq
\imag \zeta'_H (i\frac x2+\frac 12;-1)=\frac 1{8\pi} \left[ {\mathrm{Li}}_2 (-e^{\pi
x})-{\mathrm{Li}}_2 (-e^{-\pi x})\right],
\eeq
and then
\begin{align}
\frac{1}{2}\imag  \zeta'_k (0) = & \frac{{\cal T}}{2\pi}\left\{ -
\ee \log \left( 2 \cosh[2 \pi \sqrt{\Delta}] -2 \cosh [2\pi \ee] \right)\right. \\ \nonumber
&\left. - \frac{1}{4\pi} \left[- {\mathrm{Li}}_2 (-\exp [-2\pi (\sqrt{\Delta}+eE)])
+ {\mathrm{Li}}_2 (-\exp [2\pi (\sqrt{\Delta}+eE)])\right. \right. \\ \nonumber
&\left. \left. - {\mathrm{Li}}_2 (-\exp [2\pi (\sqrt{\Delta}-eE)])+ {\mathrm{Li}}_2 (-\exp [-2\pi
(\sqrt{\Delta}-eE)]) \right]\right\}.
\end{align}
Also in the Nariai case, the result coincides with the one obtained in the 
transmission coefficient approach. 
The calculation in the $\zeta$-function approach is much more difficult than the one in the 
transmission coefficient approach. Still, it furnishes the complete 1-loop effective action, 
and not simply its imaginary part. 

\subsection{Instability of the thermal state}

Also in this case, according to the general discussion of section \ref{vac-inst}, we get
\beqnl
<N_l^{out}>_{\beta_h} &=& \frac{1}{1+\exp [2\pi (\omega-\varphi^+)]}\cr
&+&  \frac{\cosh [2\pi eE]+\cosh [2 \pi \omega]}{\cosh [2 \pi \sqrt \Delta]+\cosh [2
\pi \omega]} \frac 12 \left( \tanh [\pi (\omega-\varphi^+)]+
\tanh [\pi (|\omega|+\varphi^-)] \right),  
\label{inst-nariai}
\eeqnl
with $\varphi^+ = 2 \ee=\varphi^-$. 
Thermality of the state affects the pair production induced by the presence of an electrostatic field, 
which is associated with the second term in  (\ref{inst-nariai}). We recall that in terms of physical 
(dimensionful) variables, by taking into account that 
$T_h = \frac{\hbar c \sqrt{A}}{2\pi k_b}$, and that $\omega_{phys}=\sqrt{A} \omega$, 
in such a way that $\beta_{phys} \omega_{phys}=2 \pi \omega$.

\section{Conclusions}
\label{conclusions}

We have studied the spontaneous emission of charged Dirac particles by three special 
dS black hole solutions which share the relevant property to allow exact calculations 
and to present spherosymmetric metrics with a two-dimensional spherical part completely 
factorized (with at most a constant warping factor). As a consequence of the latter 
feature, the problem allows a reduction {\sl \'a la} Kaluza-Klein to a two dimensional effective theory. 
This fact reflects itself in the common structure of our results in the $\zeta$-function approach 
as well as in the transmission coefficient approach, with the 4D imaginary part of the effective action 
appearing as a sum over K-K modes of 2D terms. As to vacuum instability, our double check by the aforementioned 
approaches leads to identical results. Analogous results occur 
in the case of scalar fields on the same backgrounds 
\cite{belcadallascal}. Moreover, both the ultracold I case and the 
Nariai case make evident that calculations of the imaginary part of the effective action are 
easier in the transmission coefficient approach; still, it is to be pointed out that, a priori, 
the $\zeta$-function approach furnishes the complete 1-loop effective action (and then also 
vacuum polarization effects are taken into account \cite{blissepf}), and not only an evaluation of the 
vacuum instability (which is the only outcome of the transmission coefficient approach). It is also 
remarkable that the analysis of ultracold I and the Nariai cases points out that different regions in $\omega$  
must be taken into account in the two approaches: in the transmission coefficient approach, 
only the level-crossing region is involved, instead in the $\zeta$-function approach the whole 
spectrum must be included.\\ 
As expected, the presence of an electrostatic field associated with the black hole charge 
induces a non-zero imaginary part of the effective action, i.e. to a vacuum instability 
which gives rise to the emission of charged pairs by the black hole, with a consequent discharge.\\
The ultracold II case is to some extent trivial, because it substantially reduces to  
the instability of the vacuum in a 2D Minkowski space with an homogeneous electrostatic field. 
This fact appears clearly by comparing our results e.g. with the ones for the 2D case in \cite{lin}. 
Nevertheless, the results obtained are meaningful, as we deal with the instability of 
a black hole with zero temperature.\\
The ultracold I case presents a less trivial structure, a non-homogeneous electrostatic field coupled 
with a Rindler-like horizon, and a non-zero temperature. Instability of the Boulware-like vacuum 
has been shown, and moreover also the instability of the Hartle-Hawking thermal state has been 
displayed; general calculations carried out in section \ref{vac-inst} show that the 
transmission coefficient which signals vacuum instability still plays a relevant role 
for a thermal state at finite temperature, and Thermofield Dynamics is in agreement with the 
results in \cite{kimleeyoon}.\\
The Nariai case is the most interesting one and also the most difficult, because it represents 
a charged black hole characterized by a finite spatial section and with an electrostatic field 
which reach a maximum between the black hole horizon and the cosmological event horizon (where 
it vanishes), and then in this respect it is really different from Reissner-Nordstr\"{o}m 
black holes (where the electrostatic field is mostly intense at the black hole horizon). 
Instability is again shown to occur.\\
For the given manifolds, a comparison between the exact results for the transmission coefficient 
and the WKB approximation ones obtained in \cite{belcaccia-rnds} has also been performed, 
and, as expected, we have found that WKB is exact in the ultracold II case.\\


\end{document}